\documentclass[12pt,preprint]{aastex}

\usepackage{graphicx}
\usepackage{amssymb}
\usepackage{longtable}
\usepackage{mathptmx}
\usepackage{natbib}

\newcommand{\chandra}{{\it Chandra\/}}
\newcommand{\asca}{{\it ASCA\/}}
\newcommand{\rosat}{{\it ROSAT\/}}

\newcommand{\xmm}{{\it XMM-Newton\/}}
\newcommand{\einstein}{{\it Einstein\/}}
\newcommand{\hst}{{\it HST\/}}
\newcommand{\lum}{\thinspace\hbox{$\hbox{erg}\thinspace\hbox{s}^{-1}$}}
\newcommand{\flux}{\thinspace\hbox{$\hbox{erg}\thinspace\hbox{cm}^{-2}\thinspace\hbox{s}^{-1}$}}
\begin{document}

\def\spose#1{\hbox to 0pt{#1\hss}}
\def\laeq{\mathrel{\spose{\lower 3pt\hbox{$\mathchar"218$}}
     \raise 2.0pt\hbox{$\mathchar"13C$}}}
\def\gaeq{\mathrel{\spose{\lower 3pt\hbox{$\mathchar"218$}}
     \raise 2.0pt\hbox{$\mathchar"13E$}}}

\title{HIGH RESOLUTION X-RAY IMAGING OF THE CENTER OF IC342}

\author{Daisy S.Y.~Mak, Chun.S.J.~Pun\footnote{corresponding email: jcspun@hkucc.hku.hk}}
\affil{Department of Physics, University of Hong Kong, Pokfulam Road, Hong Kong, PR China}
\author{Albert K.H.~Kong}
\affil{Institute of Astronomy and Department of Physics, National Tsing Hua University, 101, Section 2, Kuang-Fu Road, Hsinchu, Taiwan 30013, R.O.C.}
\affil{Kavli Institute for Astrophysics and Space Research, Massachusetts Institute of Technology, Cambridge, MA 02139}

\begin{abstract}

We present the results of a high resolution ($\theta_{\rm FWHM}\sim0.5''$) 12 ksec \chandra\ HRC-I observation of the starburst galaxy IC342 taken on 2 April 2006. We identify 23 X-ray sources within the central $30'\times30'$ region of IC342 and resolve the historical ultraluminous X-ray source (ULX), X3, near the nucleus into 2 sources, namely C12 and C13. The brighter source C12, with $L_{\rm 0.08-10~keV}=(6.66\pm0.45)\times10^{38}\lum$, is spatially extended ($\approx82\rm~pc\times127\rm~pc$). From the astrometric registration of the X-ray image, C12 is at R.A. = 03h:46m:48.43s, decl. = +68d05m47.45s, and is closer to the nucleus than C13. Thus we conclude that source is not an ULX and must instead be associated with the nucleus. The fainter source C13, with $L_{\rm 0.08-10~keV}=(5.1\pm1.4)\times10^{37}\lum$ is consistent with a point source located  at $6.51''$ (P.A.$\approx240^{\circ}$) of C12. 

We also analyzed astrometrically corrected optical {\it Hubble Space Telescope} and radio Very Large Array images; a comparison with the X-ray image showed similarities in their morphologies. Regions of star formation within the central region of IC342 are clearly visible in \hst\ H$\alpha$ image which contains 3 optical star clusters and our detected X-ray source C12. We find that the observed X-ray luminosity of C12 is very close to the predicted X-ray emission from a starburst, suggesting that the nuclear X-ray emission of IC342 is dominated by a starburst. Furthermore, we discuss the possibility of AGN emission in the nucleus of IC342, which we can not prove nor discard with our data.

\end{abstract}


\section{INTRODUCTION}

IC342 is a nearby (1.8Mpc; see~\citealt{Buta1999} for a review), almost face-on ($i = 25^{\circ}\pm3^{\circ}$;~\citealt{newton1980}) late-type Sc/Scd galaxy in the Maffei Group which is one of the closest groups to our Galaxy with well developed spiral arms. Infrared and radio observations indicated moderately strong nuclear starburst activities (\citealt{Becklin1980};~\citealt*{Rickard1984}) in IC342, whereas X-ray observations with \einstein~\citep{Fabbiano1987}, \rosat\ \citep*{Bregman1993}, and \asca\ \citep{Okada1998,Kubota2001} indicated that the majority of the X-ray emission came from a few ultraluminous X-ray sources (ULXs). ULXs are defined as off-nuclear X-ray sources whose luminosities are greater than $2.8\times10^{39}\lum$, the Eddington luminosity of a $20\ M_{\odot}$ black hole. One of the ULX identified, X3 (based on the designations from~\citealt*{Fabbiano1987}) was located near the galactic center, and was reported to be coincident with the center of IC342 at the resolutions of the \asca\ (FWHM$>1'$) and \rosat\ (FWHM$\approx5''$) data. 

Optical and near infrared observations of the inner region of IC342 revealed the presence of a nuclear star cluster~\citep*{Boker1999}. Photometrically distinct, luminous, and compact stellar clusters were often found in the centers of late-type spiral galaxies, such as NGC6240~\citep{lira2002} and NGC1808~\citep{Jimenez2005}. \citet*{Schinnerer2003} studied high spatial resolution (FWHM$\approx1.2'')$ millimeter interferometric observations of the $^{12}\rm CO(2-1)$ emission of IC342 and concluded that gaseous matter were accumulating at the nucleus through streaming motions. This suggested that star cluster formation could be a repetitive process and this process could have provided the fuel for starburst activities. As a result, if X3 were indeed associated with the galactic center, its emission may not be from a single ULX, but could be associated with the nuclear star cluster. However, the early identification of X3 as an ULX was uncertain due to limited resolution of the early X-ray data. 

The higher light collecting power and lower background level of \xmm\ provided a clearer picture of the nuclear X-ray source X3. A 10~ksec observation of IC342 was taken with the PN-CCD camera and two MOS-CCD cameras using the medium filter on 2001 February 11. Two independent studies by \citet{Kong2003} (hereafter K03) and~\citet*{Bauer2003} (hereafter B03) on the data set yielded similar results, with a total of about 35 sources detected. The slope of the X-ray luminosity function was determined to be $\sim0.5$, which was consistent with other starburst galaxies and Galactic High Mass X-ray Binaries (HMXBs). However, there were differences in the two studies about the nature and properties of the nuclear source. K03 found that the nuclear X-ray source was consistent with the instrumental PSF of a point source located $\sim3\arcsec$ from the galactic center, suggesting a off-center scenario. K03 fit the nuclear source with a power-law + blackbody spectral model ($kT=0.11$ keV, $N_{\rm H}=8.7\times10^{21}$cm$^{-2}$, photon index $\alpha=2$) and found the disk temperature was about 5 to 10 times lower than expected. Such cool disk system might be evidence of IMBHs (\citealp{Miller2003};~\citealp*{Miller2004}). However, we could not ignore the scenario of a stellar mass black hole with inner disk outflows during high accretion rate period~\citep{King2003}, and that of magnetic corona atop a cool accretion disk in which X-rays were reprocessed to give soft photons~\citep{Beloborodov1999,Miller2006}. The black hole mass of the nuclear X-ray source was estimated to be between 220~$M_{\odot}$ and 15000~$M_{\odot}$ (K03). Together with the unphysical spectral fit with the broken power-law model, K03 ruled out the possibility of an accreting stellar mass black hole with beamed relativistic jet emission. On the other hand, B03 suggested that the nuclear source was extended, mainly in the soft band, and its emission could be fit by a model of a point source plus a uniform disk with a radius of $\approx8''$, which was consistent with a supernova heated hot gas. Moreover, the nuclear source spectrum was fit by B03 with a MEKAL + power-law model ($kT=0.3$ keV, $\alpha=2.52$, $Z=6.39Z_{\odot}$, $N_{\rm H}=6.8\times10^{21}$cm$^{-2}$), including contribution from the Fe L-shell emission complexes at $\approx0.8$ keV and $\approx1.0$ keV. They found that a large fraction of the X-ray emission was likely to be thermal in origin. The relative contribution of the MEKAL component to the total absorbed flux was about 45\% and 1\% in the 0.5--2 keV and 2--10 keV band respectively, leading B03 to suggest that the hard component might be due to a number of circumnuclear X-ray binaries. 

The limited astrometric accuracy and spatial resolution of the previous X-ray observations restricted our ability to determine the exact nature of the X-ray properties at the center of IC342. In this paper, we report a new high resolution (FWHM $\approx0.5''$) \chandra\ {\it X-ray Observatory} observation of the starburst galaxy IC342. The observation and the data reduction procedures of the X-ray data, and the associated optical data for comparison, are presented in Section~\ref{s:obsdatared}. In Section~\ref{s:analysis}, the analysis of the data, including source detection, photometry, and astrometric registration of the X-ray sources are described. In Section~\ref{s:nuclearsource}, we present the results of the associated counterparts of the nuclear X-ray sources in optical and radio images and the properties of the nuclear sources. We discuss the nature of nuclear X-ray emission of IC342 in Section~\ref{s:nuclear} and we present a summary of our findings in Section~\ref{s:summary}.

\section{OBSERVATIONS AND DATA REDUCTION}
\label{s:obsdatared}
\subsection{X-ray}
\begin{figure}[htbp]
	\begin{center}  
	\leavevmode 
	\parbox{16cm}{  
	\includegraphics[width=16cm]{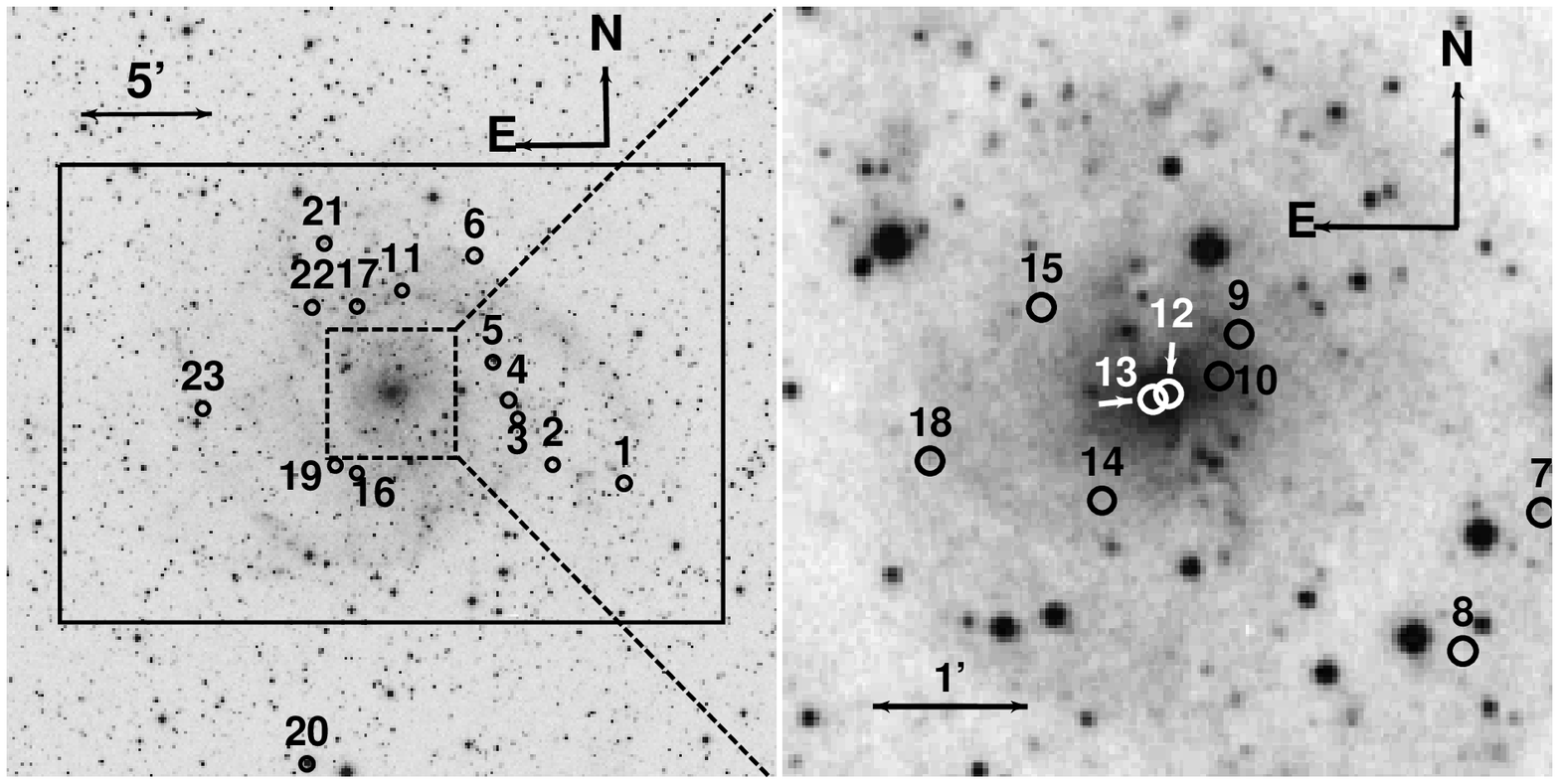} 
	\hspace*{3.5cm}(a)
	\hspace*{7cm}(b)}
\end{center}
	\caption{Digitized Sky Survey (DSS) blue band image of the field of view ($30'\times30'$) of \chandra\ (left) and the central $5'\times5'$ (right) of IC342, with detected \chandra\ X-ray sources overlaid . The radius of the source circles is $15''$ (left image) and $5''$ (right image). The dotted-line rectangle at the center in the left image showed the region enlarged in the right image. The field of view of the \xmm\ observation ($26'\times18'$) was also shown as the solid rectangular region in the left.}
	\label{f:dss}
\end{figure}

IC342 was observed on 2006 April 2 starting at 15:15:44 (UT) with the \chandra\ High Resolution Camera (HRC-I) for a total exposure time of 12.16 ksec, with a photon energy range of 0.08~-- 10~keV. The nominal pointing of the instrument was near the center of the galaxy and the total region covered by the observation was $30'\times30'$ which covered most of structure of the galaxy, including the spiral arms. The field of view of the \chandra\ observation is presented in Figure~\ref{f:dss} (left), overlaid with the Digital Sky Survey (DSS) image of IC342. The central $5'\times5'$ of IC342 is magnified in Figure~\ref{f:dss} (right) for reference. The X-ray data was analyzed using the Chandra Interactive Analysis of Observations software package 
(CIAO) v.3.3 and  Chandra Calibration Database
(CALDB) v.3.2.1 software applied on the level-2 event fits file. The \chandra\ data had gone through Standard Data Processing\footnote{http://cxc.harvard.edu/ciao/dictionary/sdp.html} (SDP, or the pipeline) with updated calibration procedures, including instrumental corrections and filtering good time intervals, applied. 

There were 4 other archival \chandra\ IC342 data sets: three AXAF CCD Imaging Spectrometer (ACIS-S) observations (each 10 ksec) in sub-array mode centered on one of the ULX (X1), and one 3 ksec HRC-I observation. For the 3 ACIS-S data, the detectors did not cover the nuclear region of IC342 as the sub-array mode was used to reduce pile-up. The 3 ksec HRC-I observation, on the other hand, had insufficient signal for source detection at the galactic center. Therefore only the 12 ksec HRC-I image was analyzed in this work.

IC342 was also observed 4 times between 2001 and 2005 with \xmm. Among these 4 observations, the data taken in year 2001 were studied by B03 and K03. We have retrieved the 2001 EPIC data for the analysis of X-ray spectrum of the nuclear sources.  The event files were reprocessed and filtered with the \xmm\ Science Analysis Software (SAS v7.1.0) and we used the HEAsoft v6.2 and XSPEC v12.3.1 packages to perform spectral analysis. 

\subsection{Optical}
We also analyzed archival {\it Hubble Space Telescope\/} (\hst)\ broadband and emission-line images of IC342 for comparisons with the X-ray data. We searched the \hst\ Data Archive for images with pointings that included the central part of the galaxy. The calibrated and data-quality images were retrieved, with preliminary CCD processing, including bias and dark subtraction, flat-fielding, and shutter-shading correction, automatically applied using calibration frames closest to the time of the observations. The fields were inspected to ensure that the galactic nucleus did not fall on the boundaries on any of the CCD chips. We found that most of the archival images were saturated at the central region of the galaxy. The only unsaturated images were the Wide Field Planetary Camera 2 (WFPC2) images taken on 1996 January 7 in broadband {\it V} (F555W), {\it I} (F814W), and in narrow-band H$\alpha$~(F656N), and they were selected for comparison with our \chandra\ HRC-I data. The exposure time of each frame was 260 sec each for {\it V} and {\it B} band and 500 sec for H$\alpha$. Mosaics were produced using the STSDAS task WMOSAIC which removed known relative rotations and geometric distortions of the individual WPFC2 chips. Astrometry of the calibrated \hst\ images were corrected using the coordinates from the USNO-B1.0 Catalog~\citep{Monet2003} for comparison with the positions of X-ray sources in the same reference frame. Details of the astrometric registration of the data will be described in Section~\ref{s:anlyast}.
 
\section{ANALYSIS}
\label{s:analysis}
\subsection{X-ray Source Detection}
\label{s:anlydetect}
We performed source detection on the X-ray image using the WAVDETECT algorithm~\citep{Freeman2002} which employs a wavelet method to detect point sources. This algorithm was selected because it is able to effectively identify extended sources with properly chosen wavelet scales; this is important for the analysis of our \chandra\ IC342 image, which has an extended source near the galactic center. Source detection results from the WAVDETECT algorithm depend on two parameters, the significance threshold~({\it sigthresh}) and the image scale. We set the wavelet radii, in pixels, to be `1.0 1.414 2.0 2.828 4.0 5.657 8.0 11.314 16.0' and the parameter {\it sigthresh} to be $2\times10^{-7}$ for all runs. This detection threshold corresponds to less than one false detection in the \chandra\ HRC-I field due to statistical fluctuations in the background~\citep{Freeman2002}. 

The runtime of WAVDETECT depends heavily on the selected resolution of the image, and a resolution of $1024\times1024$ was considered to be the practical size limit for the input data on most desktop computers. We tested running WAVEDETECT on the central part of the image containing the galactic center with binning factors 1, 2, 4, 8, 16, and 32. An exposure map, created using CIAO, was applied for each binning to normalize the images and to correct for  instrumental artifacts caused by imperfections of the \chandra\ mirror and detector, especially at the edges of the field. We found the list of sources detected  depended on the binning factor applied. A binning factor of 4 was used in the final data reduction as a compromise between spatial resolution and faint source detection. The \chandra\ field was divided into 4 equal quadrants with overlapping at the central regions before being recombined to the full-frame. The final  image scale with a binning factor of 4 was $0.53''$/pixel. 

To generate the final source list, we applied two additional selection criteria for the sources identified by the WAVDETECT algorithm:
\begin{enumerate}
 \item The source should be detected in at least 2 different binning images, with one of them being the one with a binning factor of 4. 
 \item The S/N of the source had to be greater than 3 in the binning factor 4 image.
\end{enumerate}
With these selection criteria we detected a total of 23 X-ray sources in our \chandra\ HRC-I IC342 image. Properties of the detected sources are summarized in Table~\ref{t:sourcelist}. A total of 14 and 8 of our sources were also previously detected in the \xmm\ (B03; K03) and the \rosat\ observations~\citep{Bregman1993} respectively. The alternate \xmm\ and \rosat\ source IDs are listed in Table~\ref{t:sourcelist}. The locations of all sources detected are shown overlaid on the DSS image in Figure~\ref{f:dss}. A majority of the identified sources are located in the spiral arms of the galaxy, as previously noted by B03 and K03 with \xmm\ observations. Using the Chandra Deep Field data~\citep{Brandt2001}, we estimated that there could be 9--10 background objects within our $30'\times30'$ \chandra\ frame, implying that $\sim~40$\% of the sources currently detected could be background objects. 

\begin{figure}[htbp]
	\begin{center}
	\leavevmode
		\parbox{12cm}{
	\includegraphics[width=12cm]{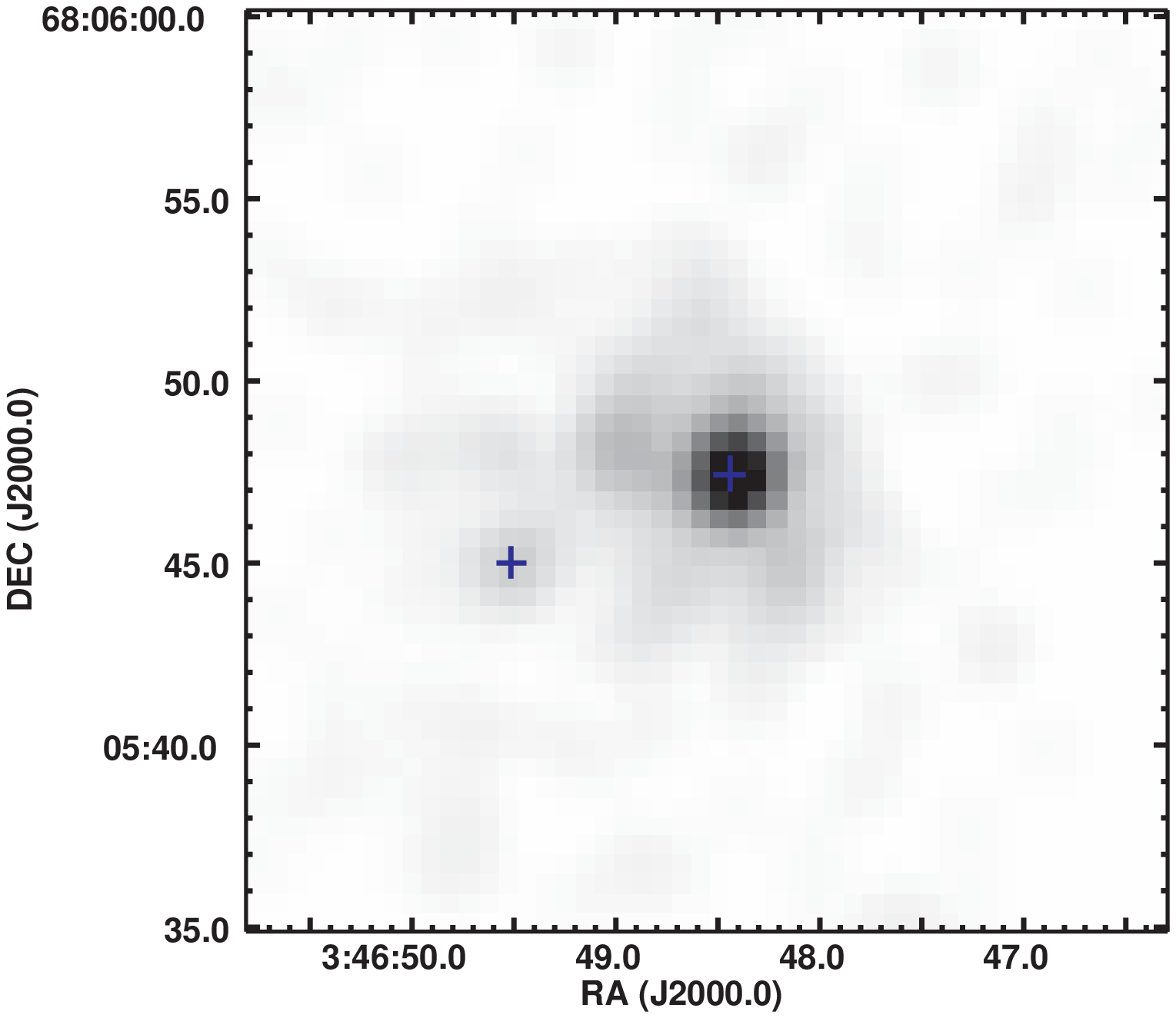}} \\
		\vspace*{0.005in}
		\parbox{12cm}{
	\includegraphics[width=12cm]{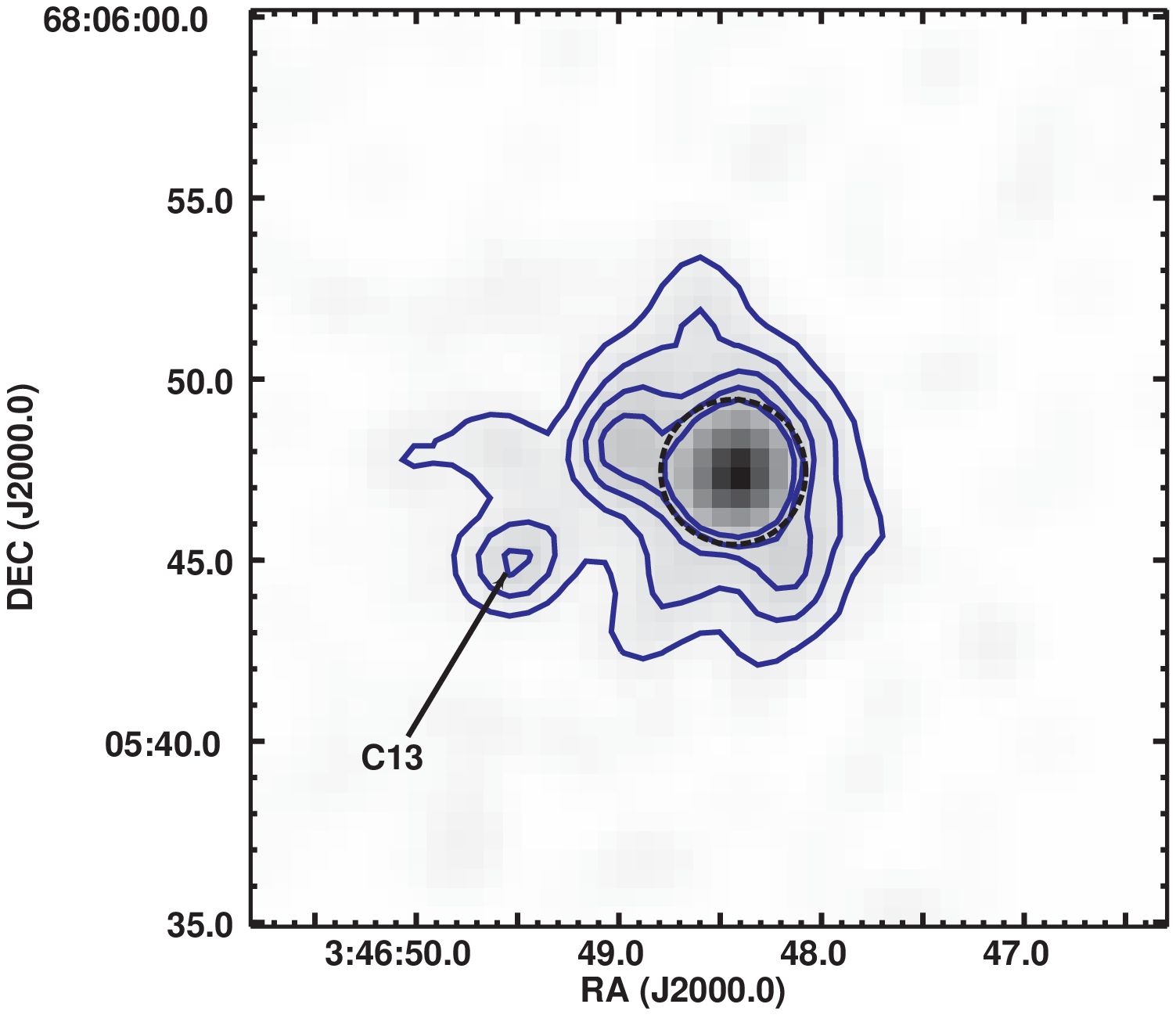} }\\
	\end{center}
	\caption{{\tiny \textbf{Top}: The smoothed \chandra\ HRC-I image of central $25''\times25''$ region of IC342. The image was smoothed with a Gaussian kernel radius=3 and displayed with linear scale at 99.5\%. The centroid of C12 at R.A. = 03h:46m:48.43s, decl. = +68d05m47.45s is marked with a blue cross at the center of the image. The dimmer source C13 is located $6.51''$ from C12 at P.A.$\approx240^{\circ}$ and its position is also marked by a blue cross. \textbf{Bottom}: The \chandra\ HRC-I image (same as above) with X-ray contours (blue) superposed. The data have been smoothed with gaussian function, and contours are at 0.3, 0.6, 0.9, 1.2, and 1.5 counts. The dashed line circle showed a $4''$ region at the center. Details of the X-ray morphology are discussed in Section~\ref{ss:xrayimage}. } } 
	\label{f:central_xray}
\end{figure}

With the high resolution \chandra\ HRC-I image, we were able to resolve for the first time X-ray emission from the center of IC342 into two distinct components, namely C12, the brighter and extended component near the center of the galaxy, and C13, a much fainter component located $6.51''$ from C12 at P.A. $\approx240^{\circ}$. The two sources could be identified by the WAVEDETECT algorithm when the binning factor were set to be either 2 or 4, suggesting that they were genuine detections. Images of the central $25''\times25''$ of the \chandra\ HRC-I image of IC342 are shown in Figure~\ref{f:central_xray}, together with the positions of the two nuclear sources C12 and C13, and the X-ray flux contours. Details of the X-ray emission morphology will be discussed in Section~\ref{ss:xrayimage}.

\subsection{Photometry}
\label{s:anlypho}
We extracted the integrated counts of the 23 detected X-ray sources of IC342 using the CIAO tool DMEXTRACT, and listed the results in Table~\ref{t:sourcelist}. The source regions were defined in DS9 using circular apertures with different radii based on the relative brightness of the source compared to the local background and their off-axis angles. All the background regions were extracted using circular annuli centered on individual sources. The only exception was the nuclear source C12, in which an elliptical aperture was used because of its extended nature.  Moreover, the background regions of the two nuclear sources C12 and C13 ($6.51''$ apart) had to be modified to avoid overlapping.  We used WebPIMMS to convert the count rates for all detected X-ray sources into unabsorbed 0.08--10~keV luminosity by assuming an absorbed power-law model with $N_{\rm H}=8\times10^{21}$cm$^{-2}$ and $\alpha=2$, adopted from the average parameters derived from spectral fitting of the \xmm\ data by K03. These results are also presented in Table~\ref{t:sourcelist}. This set of average parameters was similar to that derived from the same \xmm\ data set by B03, with $N_{\rm H}=3.4\times10^{21}$cm$^{-2}$ and $\alpha=1.63$. The difference in 0.08--10 keV luminosities obtained with these two sets of spectral parameters is about 10\%, with slightly lower values for the parameters from K03.

For C12, B03 and K03 derived the X-ray flux by fitting the source spectrum with a simple absorbed power-law model (B03: $N_{\rm H}=4.0\times10^{21}$cm$^{-2}$ and $\alpha=2.67$; 
K03: $N_{\rm H}=3.2\times10^{21}$cm$^{-2}$ and $\alpha=2.53$). For comparison, our adopted flux for C12 is smaller by about 40\% from the B03 model and about 30\% larger than the K03 simple power-law model by using the average the X-ray sources parameters. In this paper, we adopt the value from K03, except for Table 1 in which we use the average parameters from K03 as described above.

There is a large uncertainty in the distance estimate to IC342 due to the large uncertainty in the amount of dust extinction toward the Galactic Center along that line-of-sight. Distance estimates of IC342 range from 1.5 to 8 Mpc~\citep{Buta1999}. A distance of 1.8 Mpc was assumed in this paper (corresponding to a scale of $10'' =$~87 pc), identical to that in Boker et al.~(1997, 1999) and K03, and similar to that (2~Mpc) assumed by \citet{Meier2005}. Using an elliptical ($4.6''$ and $9.9''$ for semiminor and semimajor axis) integrating window to include all the diffused emission, we found the stronger central nuclear source C12 has $255\pm17$ net counts, corresponding to an unabsorbed X-ray luminosity of 
$L_{\rm 0.08-10~keV} = (6.7\pm0.5)\times10^{38}\lum$; the weaker source C13 was one of the weakest source detected in our image, with a net count of $15\pm4$ (S/N = 3.8), 
and an unabsorbed luminosity $L_{\rm 0.08-10~keV} = (5.0\pm1.4)\times10^{37}\lum$.

\subsection{Astrometric Registration}
\label{s:anlyast}
One of the main objectives of our study is to obtain accurate positions of X-ray sources near the center of IC342. The \chandra\ HRC-I images have nominal 90\% and 99\% confidence astrometric accuracy of $0.6''$ and $0.8''$ respectively, while the astrometric accuracies of the \hst\ WFPC2 data is about $1''$~\citep{Ptak2006}. In order to identify optical counterparts of our \chandra\ X-ray sources in the \hst\ data, we first registered their positions to the same reference frame for comparison. We used the USNO-B1.0 Catalog~\citep{Monet2003} to improve the astrometric accuracies for both the \chandra\ and \hst\ images. The advantage of using this is that proper motions are already included in the catalog, and hence can be corrected for. The nominal RMS accuracy of the USNO B1.0 catalog is $0.2''$. 

The astrometry of our \chandra\ image can be improved when one or more X-ray sources are associated with optical or radio sources of known positions. As seen in Figure~\ref{f:dss}, one of the brightest sources, C5, is clearly associated with a bright stellar object within 1$''$ in the DSS image. This source was also previously identified to be associated with a star in the \rosat\ (Source 4 in~\citealt{Bregman1993}) and the \xmm\ observation (X12 in K03). The spectrum of X12 is soft and can be fitted with a blackbody model ($kT$ = 0.17 keV). B03 determined the X-ray to optical flux ratio $\log(f_{\rm x}/f_{\rm opt})$ of X12 to be $\leq$-3, which was consistent with that expected from a foreground star of $\log(f_{\rm x}/f_{\rm opt})$ $\leq$-1. B03 and K03 also identified it to be a foreground star. Using the USNO-B1.0 Catalog, we found that the source C5 was offset by $0.42''$ (RA offset by $0.13''$ and DEC offset by $0.40''$) from this foreground star, which was within the expected $0.5''$ pointing accuracy of the \chandra\ data. While there are other X-ray sources which could be associated with point sources in the DSS image, their positional offsets are either $\geq0.5''$ or their S/N ratios were too low ($\leq3$) to make firm registrations with their optical counterparts. Thus we corrected the pointing of the \chandra\ image using the CIAO command WCSUPDATE by linearly shifting it until C5 and the star coincided. This astrometric reference was applied to all the sources and their shifted positions were tabulated in Table~\ref{t:sourcelist}. 

The astrometry of the \hst\ images were corrected using the IRAF task CCMAP. With many bright stars within the WFPC2 frame, we could visually identify stars from the catalog in the \hst\ data products and compare them with the USNO-B1.0 Catalog to improve the astrometry. A total of 7 stars were used for astrometric registration for the {\it V} and {\it I} band data, and 3 stars for the H$\alpha$ data. The RMS positional errors returned by CCMAP were $0.30''$ for {\it V}, $0.32''$ for {\it I}, and $0.0021''$ for H$\alpha$. These registrations are similar to that of our \chandra\ HRC-I images, hence allowing meaningful spatial comparison of the sources in different wavelengths.

\section{RESULTS}
\label{s:nuclearsource}
\subsection{Morphology of the IC342 center}
\subsubsection{X-ray}
\label{ss:xrayimage}
The high spatial resolution of the \chandra\ HRC-I image ($\theta_{\rm FWHM}\sim0.45''$) provides an unprecedented detailed look into the X-ray emitting structure of the IC342 galaxy center.  Detailed structures of the X-ray emission from the IC342 center are illustrated in Figure~\ref{f:central_xray} with the central $25''\times25''$ of the \chandra\ image. The two nuclear sources, C12 and C13, are separated from other X-ray sources and stand out from the low diffuse X-ray background. The brighter source, C12, was observed to have at least two components. The first component is the core emission of radius $\sim2''$ at the center, roughly consistent with that bounded by the X-ray contour line of 1.4 counts. The second component is the diffuse emission out to a radius of $5''$, with an asymmetric extension along the north-south direction. The spatial extent, derived from the X-ray image at 99.5\% linear scale, is $14.6''$ in the north-south and $9.4''$ in the east-west directions (127~pc~$\times$ 82~pc assuming a distance 1.8~Mpc), resulting in a major-to-minor axis ratio of $\sim1.6$. The observed north-south elongation is similar to that observed in the near infrared wavelength by~\citet*{Boker1997}, who detected a diffused emission with a major-axis of 110 pc and a high surface brightness source at the center. The authors interpreted the results as evidence for a small-scale nuclear stellar bar and a central concentration of stars.  

We determined the X-ray luminosity of the core and the circumnuclear diffuse regions of C12 by extracting respectively the source counts of the inner $4''$ region with a circular aperture, and the source counts of an annulus of $r=2''-5''$.  The resulting X-ray luminosities (assuming the same spectral and distance parameters as listed in Section~\ref{s:anlypho}) are $L^{\rm core}_{\rm 0.08-10~keV} = 3.14\times10^{38}\lum$ and $L^{\rm diffuse}_{\rm 0.08-10~keV} = 3.34\times10^{38}\lum$, implying roughly equal flux contributions in X-ray from these two regions.  

\begin{figure}[htbp]
        \parbox{7cm}{
	\includegraphics[bb= 18 164 540 670,width=7cm, angle=0]{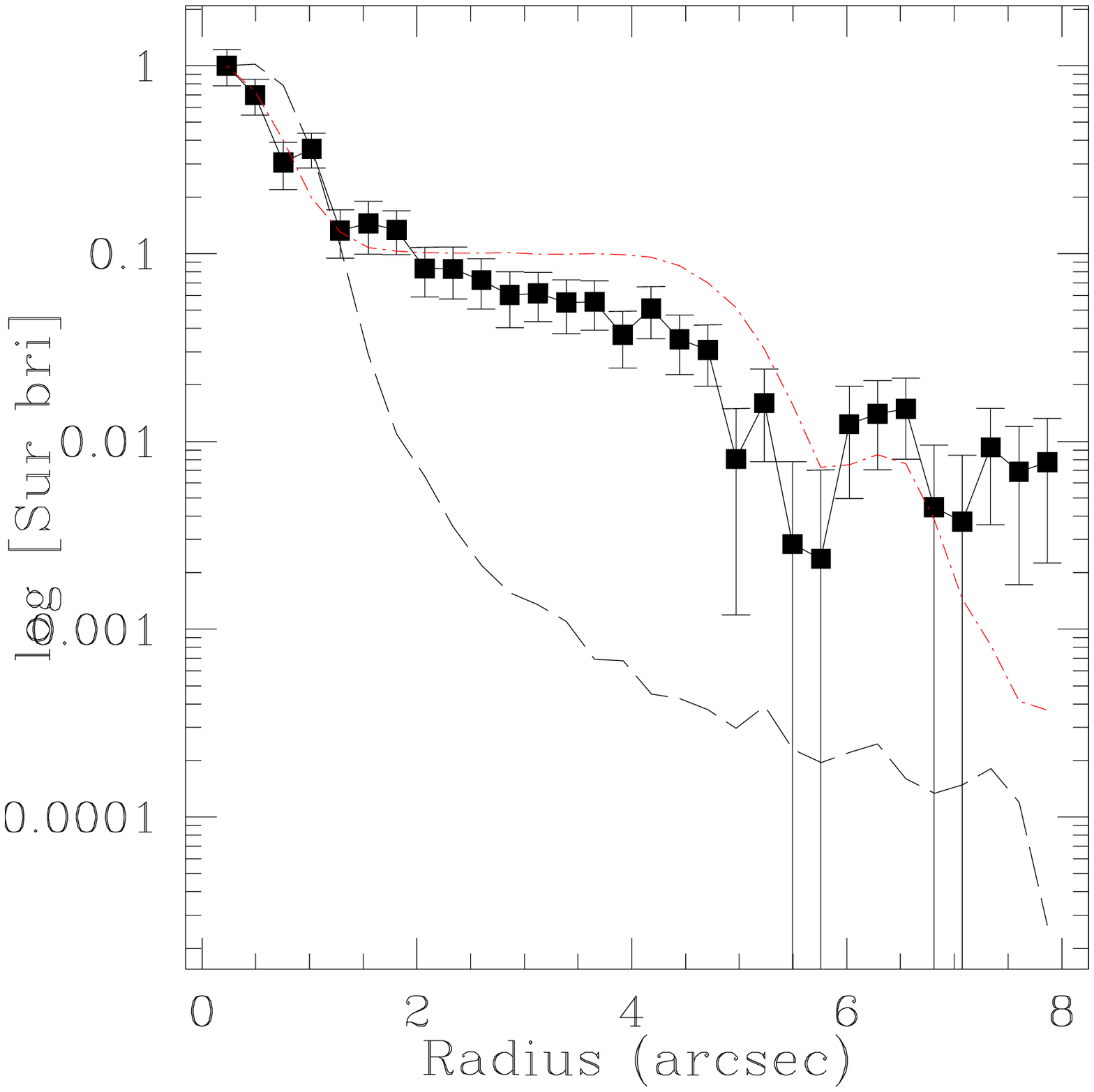}\\
	}
	\hfill
       \parbox{7cm}{
	\includegraphics[bb= 18 164 540 670,width=7cm, angle=0]{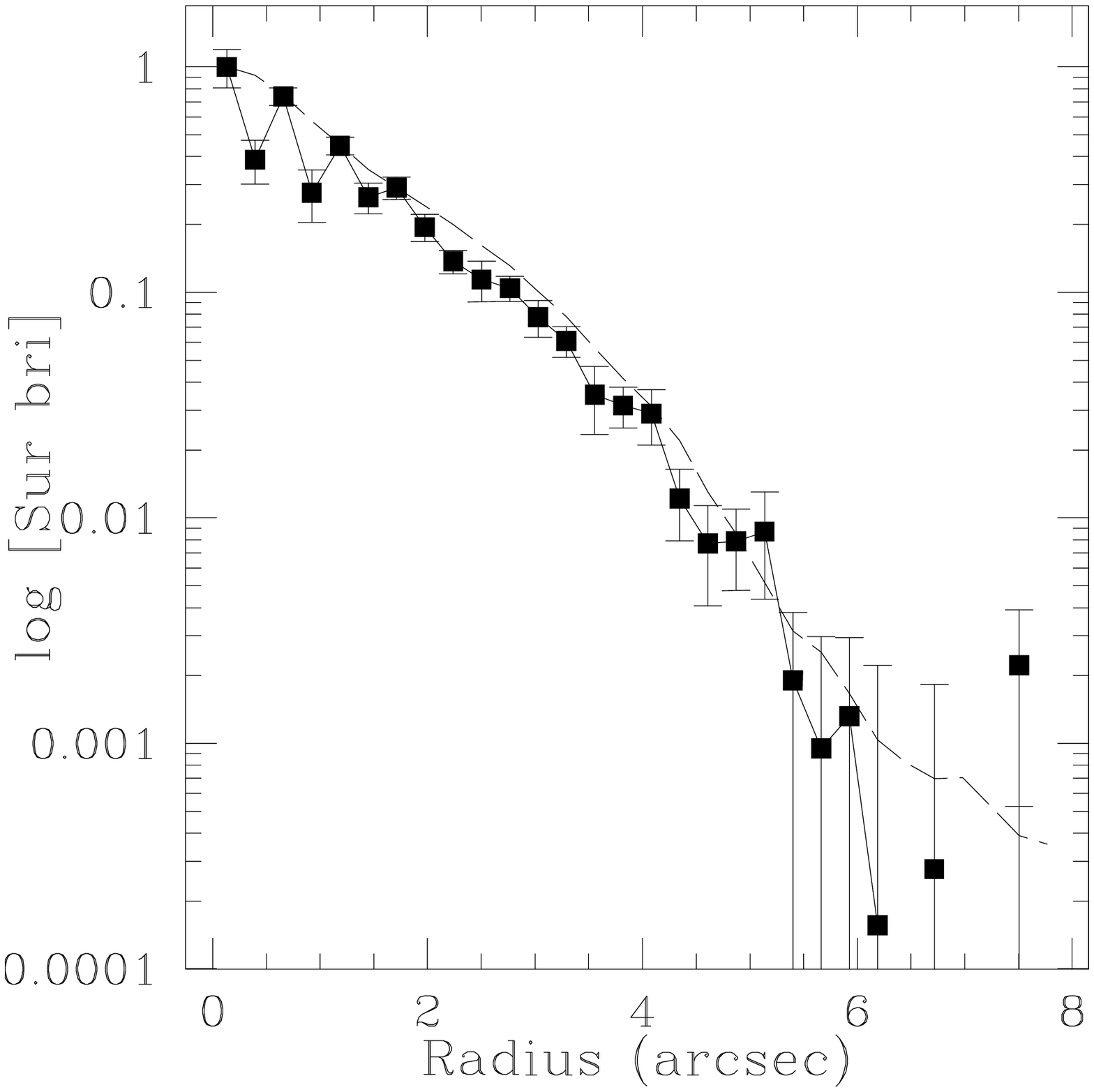}\\
	}

        \vspace{0.5cm}
        \parbox{7cm}{
	\includegraphics[bb= 18 164 540 670,width=7cm, angle=0]{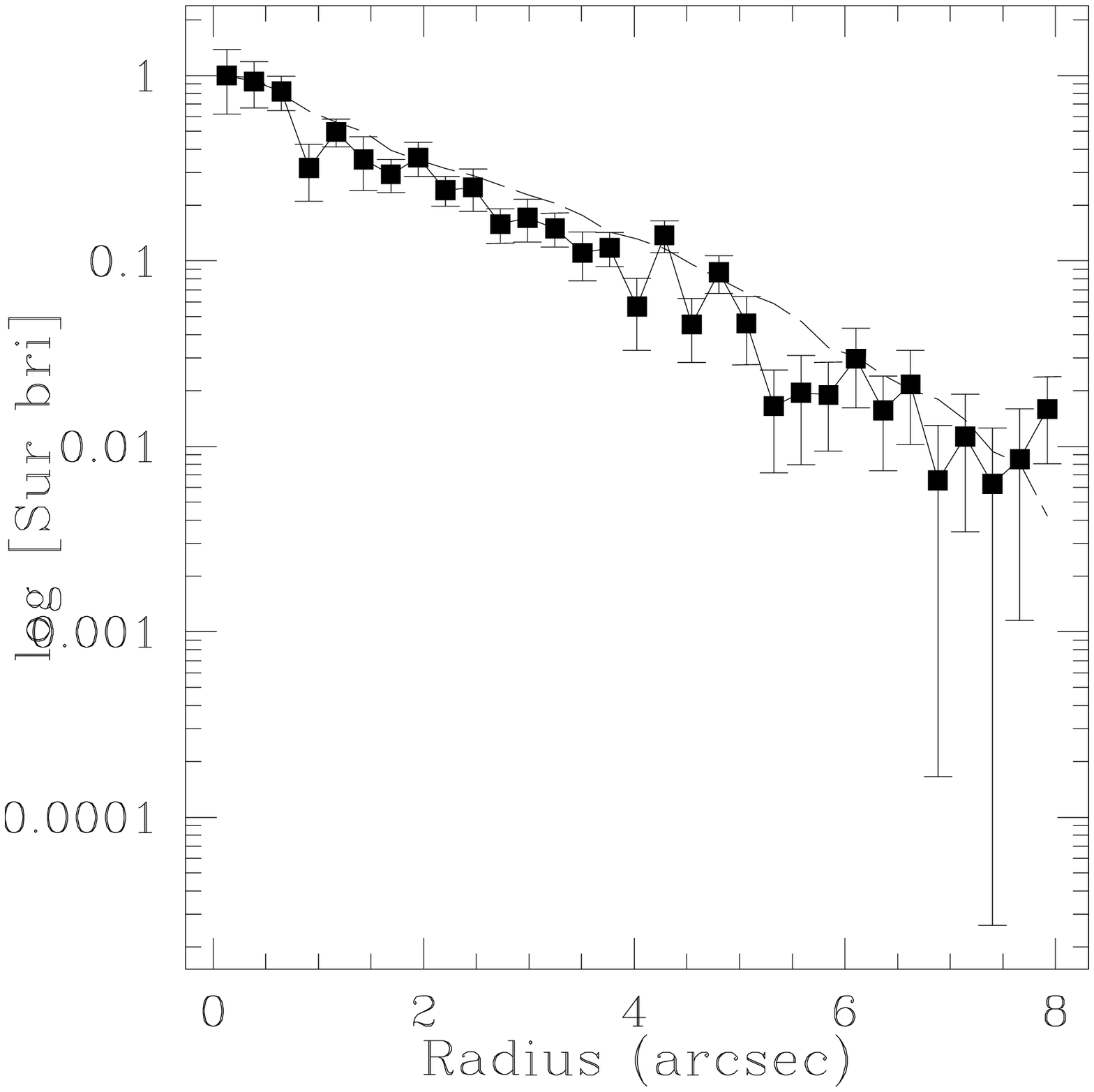}\\
	}
	\hfill
        \parbox{7cm}{
	\includegraphics[bb= 18 164 540 670,width=7cm, angle=0]{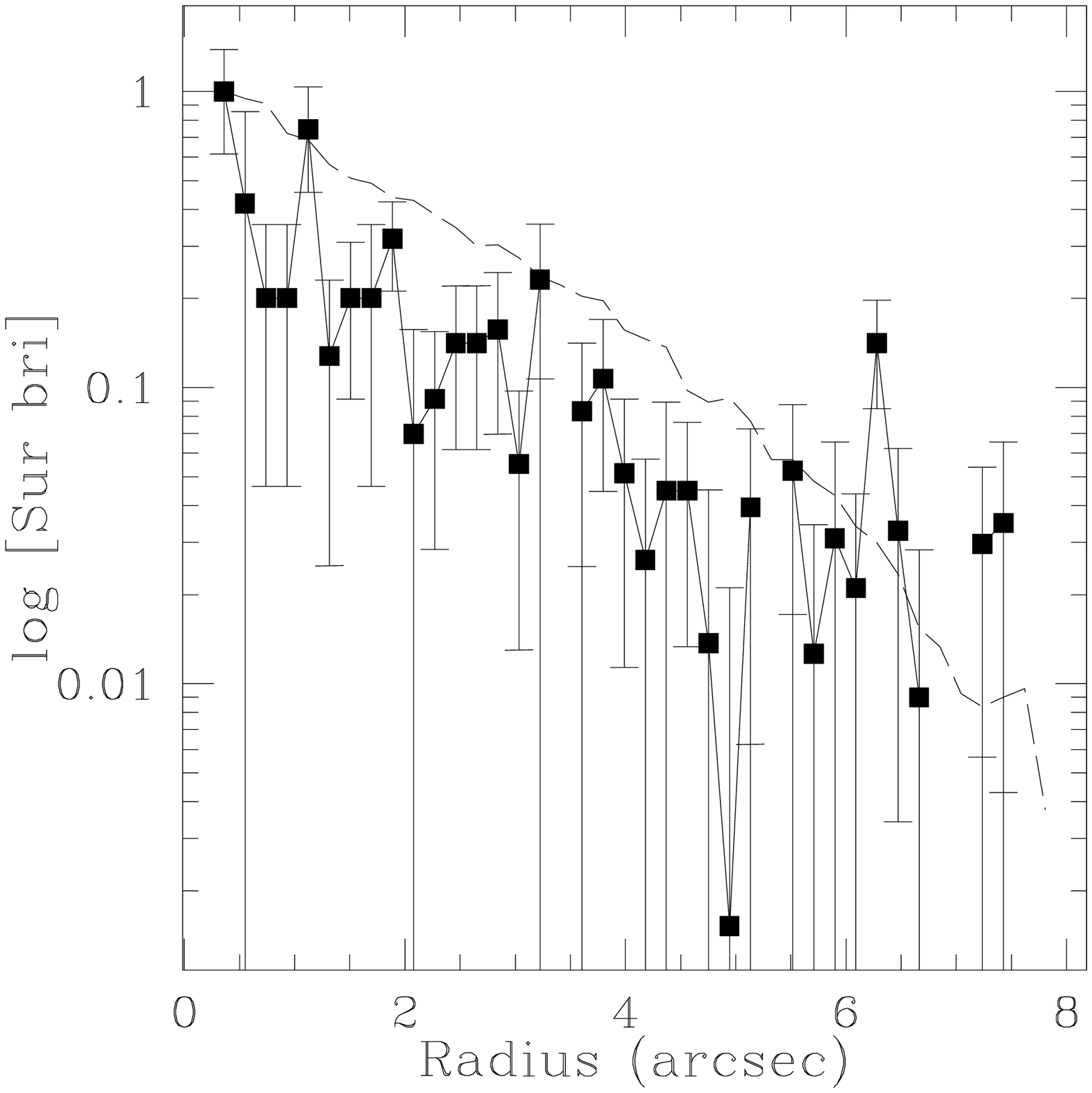}\\
	}
	\caption{The observed radial profiles of X-ray sources C12 (top-left), C3 (top-right), C6 (bottom-left), and C21 (bottom-right) (solid square) and the calculated profiles of the PSF from the ChaRT program (black dashed line) at the location of the sources. The errors on the ChaRT PSF are negligible. For the radial profile of C12, the radial profile of a model with one point source overlapped on a circular disk of $r=5''$ (red dashed line) was also shown. The bump in the radial profile of C12 between $6''$ and $7''$ is from C13. 
	\label{f:rp}}
\end{figure}

To analyze the structure of emission from C12, its radial flux profile was extracted and normalized at the center. The results are shown in Figure~\ref{f:rp}~(top-left).  As a comparison, the normalized radial flux profile of the \chandra\ PSF at the location of C12 derived from the \chandra\ PSF simulator, the \chandra\ Ray Tracer (ChaRT)\footnote{http://cxc.harvard.edu/chart/}, is also shown ({\it dash\/}).  The emission from the core component of C12 at the central $r < 2''$ is consistent with that from a point source. However, the diffused emission component at radius $r\ge2''$ does not decrease as steeply as that of the PSF.  We also computed the radial flux profile of a point source plus a uniform $r=5''$ circular disk  at the location of C12 computed from the simulation task MARX\footnote{http://space.mit.edu/ASC/MARX/}.  The results ({\it dot-dash\/}), also shown in Figure~\ref{f:rp}~(top-left), are generally consistent with that observed for C12, with a hint of slowly-decreasing surface brightness in the range $r\sim2''-5''$ for C12.  A Kolmogorov-Smirnov test (K-S test) was performed to measure the difference between radial profile of C12 and the models. The K-S test statistics $D$, the maximum vertical deviation between the two curves, is 0.78 (confidence probability$<0.01\%$) for C12 and a point source, and 0.33 (confidence probability = 6.0\%) for C12 and a point source plus a $5''$ disk.  The surface brightness profiles for selected bright X-ray sources C3, C6, and C21, are also plotted in Figure~\ref{f:rp} (top-right), \ref{f:rp} (bottom-left), and~\ref{f:rp} (bottom-right), respectively for comparison. The K-S statistic $D$ of the reference sources C3, C6, and C21 are in a much lower range of 0.17--0.33 (corresponding to confidence probabilities 2.2\%--76.0\%). These results reinforce the conclusion that the emission from C12 is not consistent with that from a point source; the X-ray emission of C12 might be due to multiple sources or diffuse emission or a combination of both. However, we cannot conclude whether the emission is consistent with the various components proposed by~\citet{Boker1997} due to the limited resolution of the X-ray data.

The newly detected X-ray source C13 is one of the faintest source in our \chandra\ observation, with $L_{\rm C13}/L_{\rm C12}\approx6\%$. 
We failed to extract the radial profile of C13, and cannot determine whether it is a point source or a diffused source because of insufficient count statistics.  We also could not exclude the possibility that the emission was actually part of the nebulosity of the extended emission component of C12.

\subsubsection{Comparison of X-ray with optical data}
\label{ss:oc}
\begin{figure}[htbp]
	\begin{center}
	\leavevmode
	\includegraphics[width=12cm]{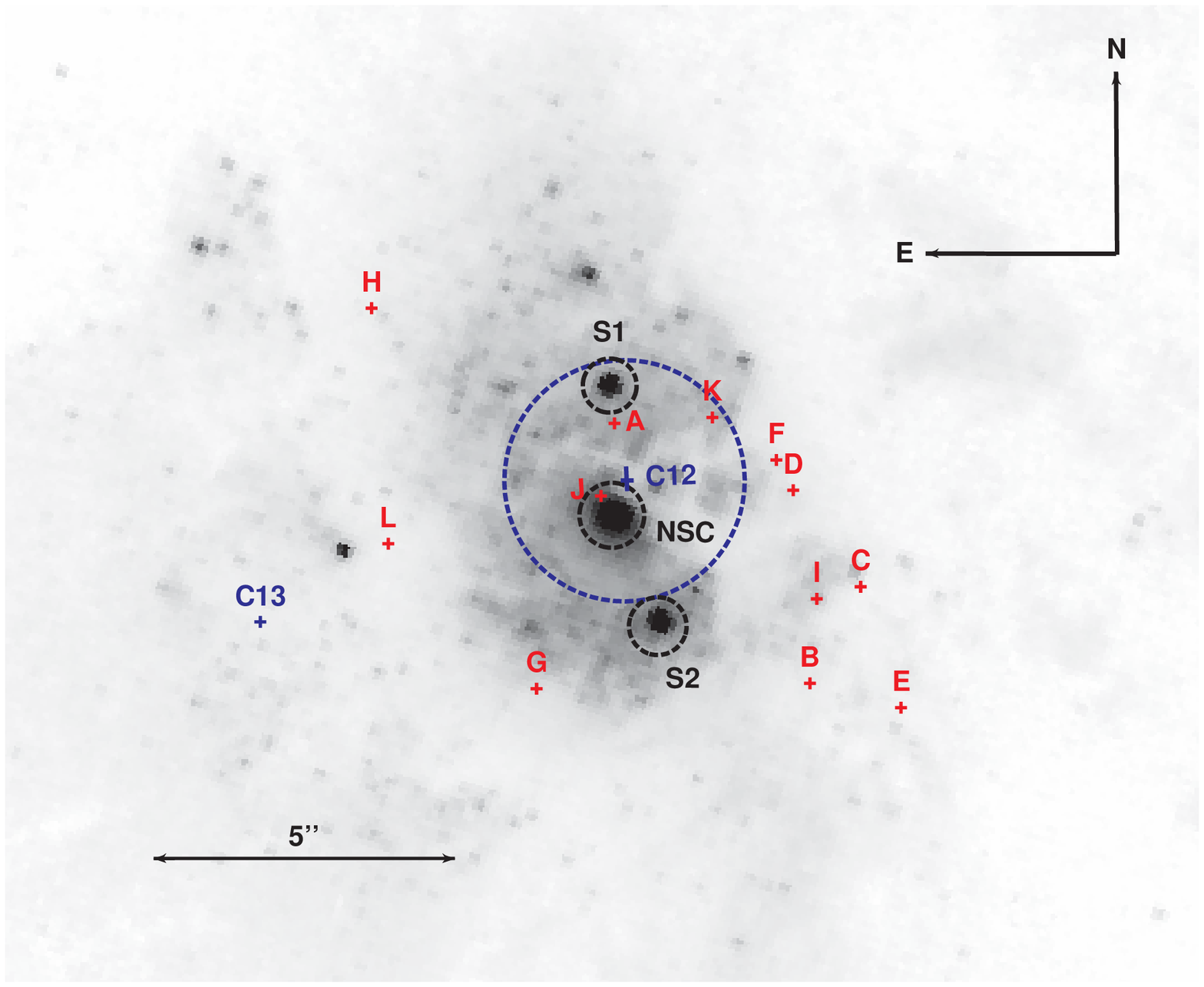} \\
	\includegraphics[width=12cm]{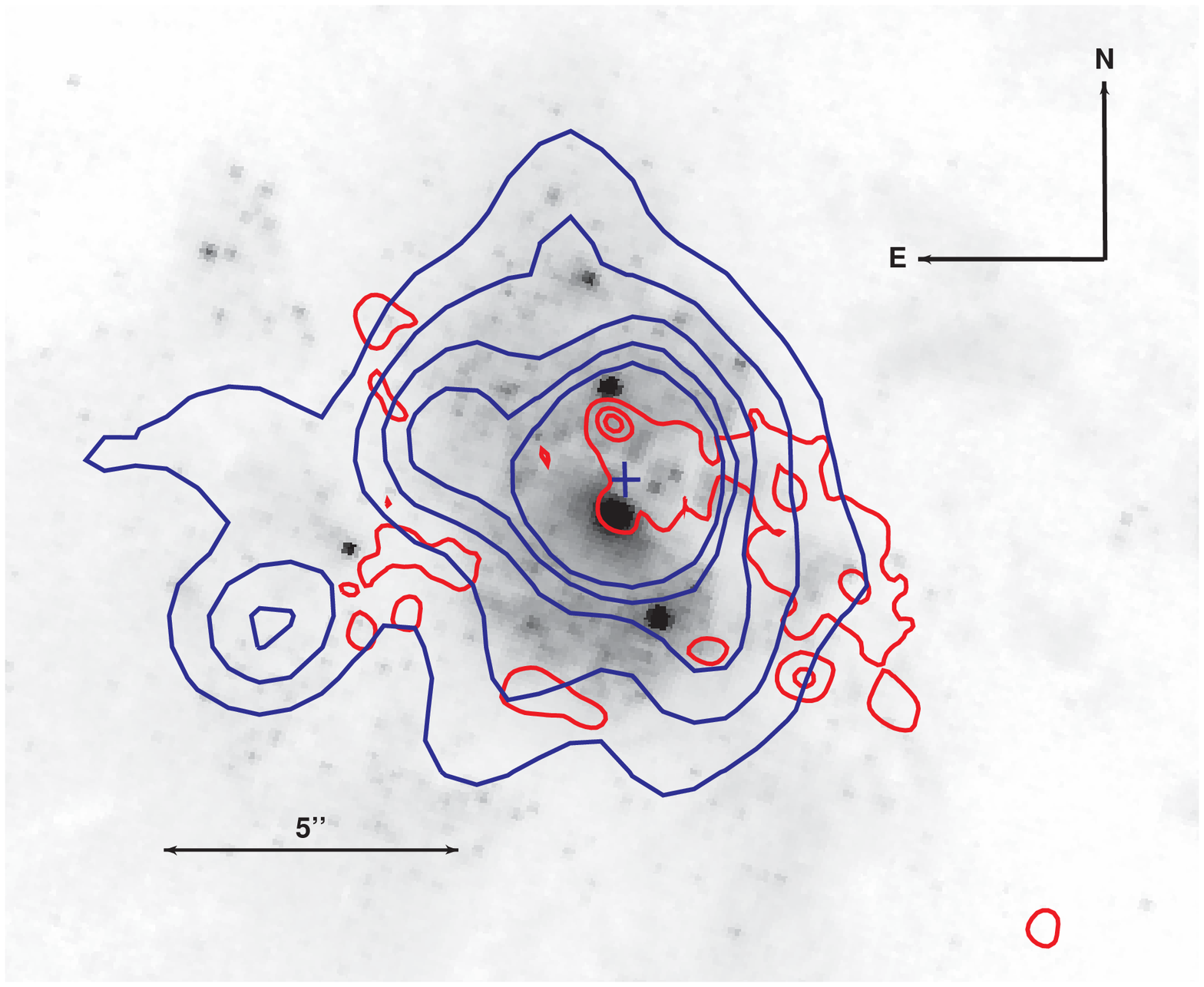} \\
	\end{center}
	\caption{{\scriptsize \hst\ {\it V} band image of the central $19''\times16''$ region of IC342 displayed in square root scale to bring out the features. \textbf{Top}: Locations of the 3 optical star clusters are labeled by black circles. The centroids of the X-ray center sources C12 and C13 (blue cross), and a $4''$ diameter region centered around of C12 (blue dashed circle), are also shown. Radio sources detected in the VLA 2 cm and 6 cm maps are marked and labeled (red cross) according to the naming convention by \citet{Tsai2006}. \textbf{Bottom}: Overlay of \chandra\ X-ray (blue; same as figure~\ref{f:central_xray}) and radio 6~cm contours (red) on the optical data. Contour levels of radio 6~cm flux are 0.37, 0.93, 1.49, 2.05, 2.61~mJy~beam$^{-1}$.} }
	\label{f:vband}
\end{figure}

\begin{figure}[htbp]
	\begin{center}
	\leavevmode
		\parbox{12cm}{
		\includegraphics[width=12cm]{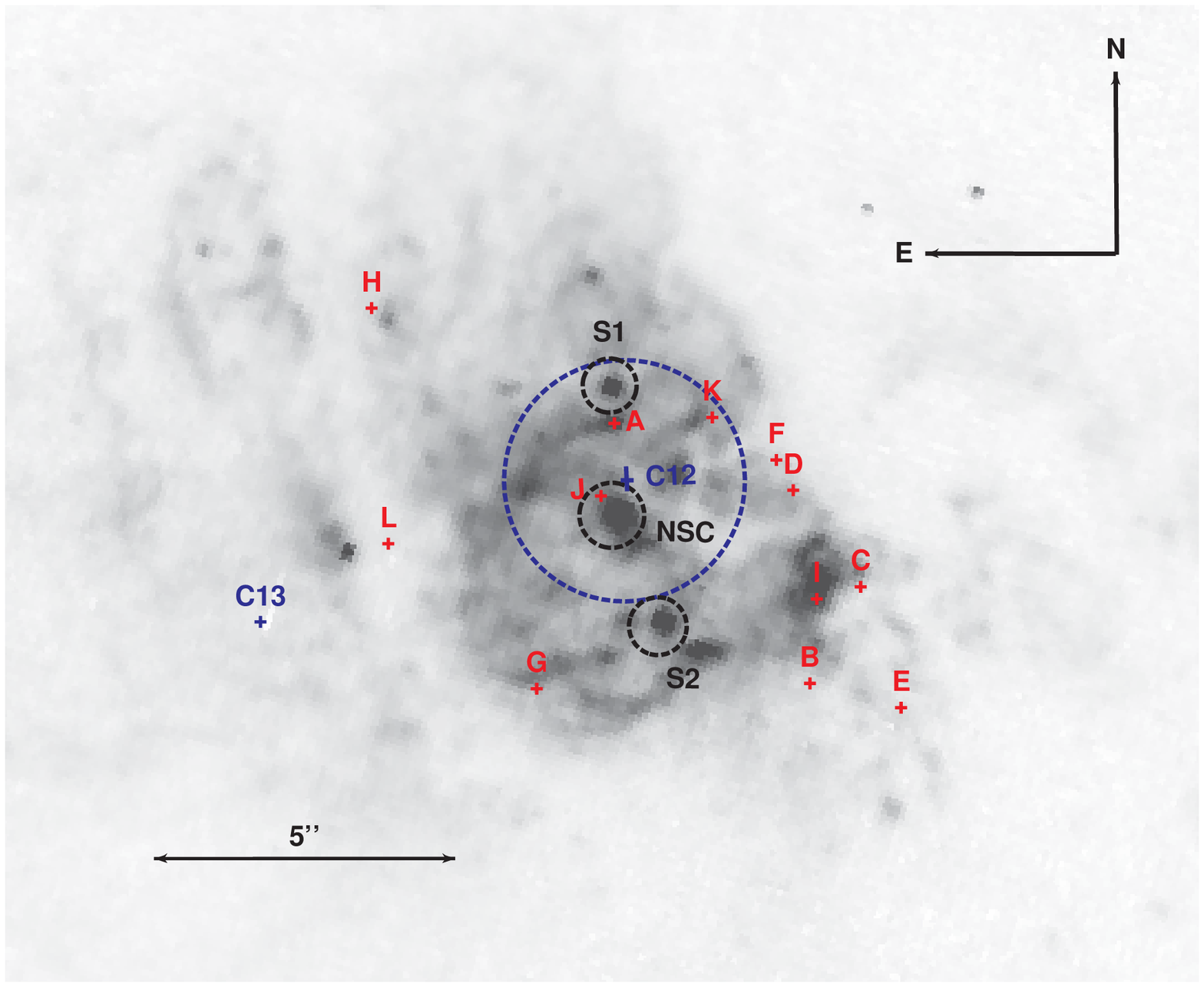} \\
		}
		\vspace*{0.005in}
		\parbox{12cm}{
	\includegraphics[width=12cm]{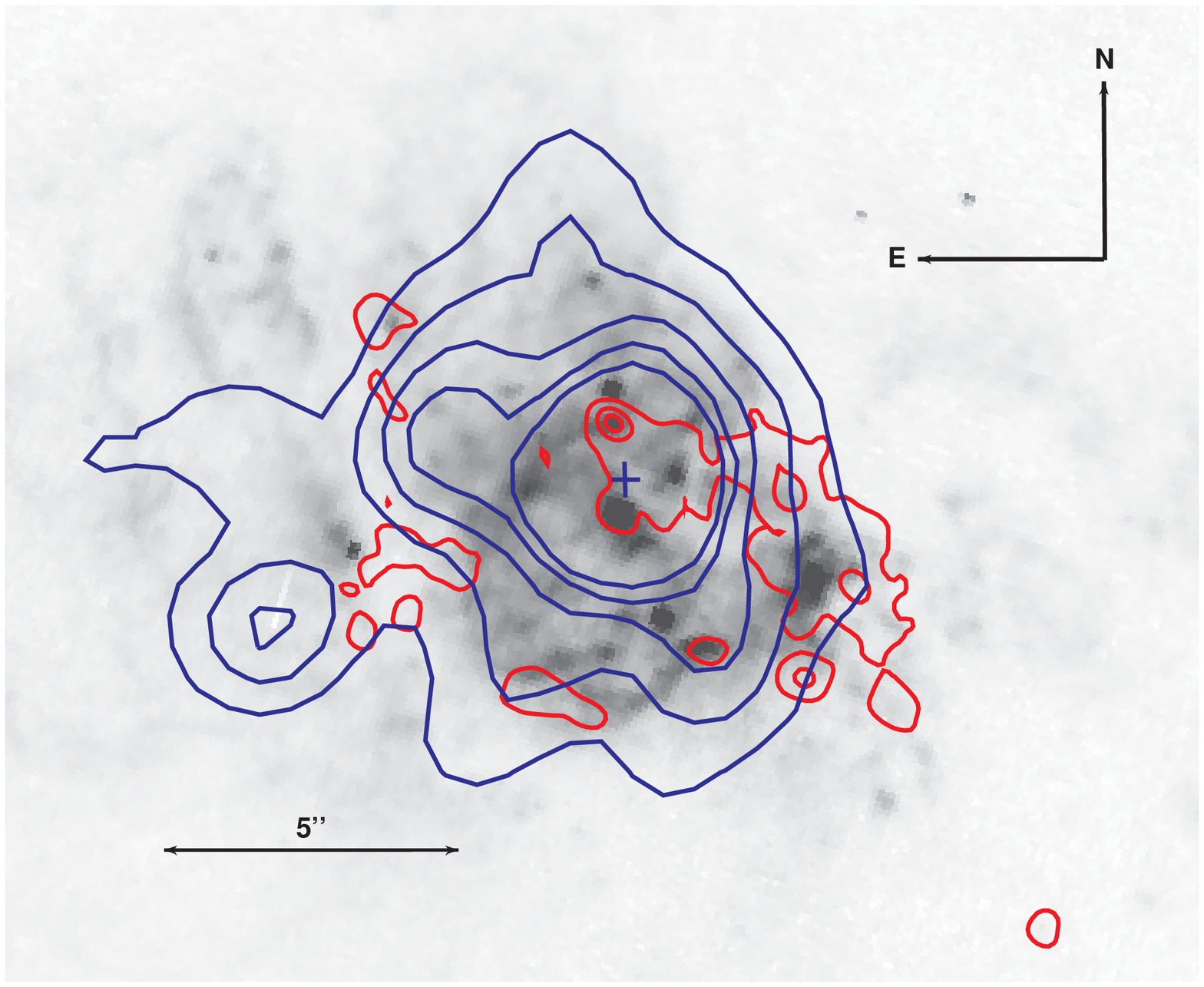} \\ 
	}
	\end{center}
	\caption{Similar to figure~\ref{f:vband}, except that the background image is~\hst\ H$\alpha$.} 
	\label{f:halpha}
\end{figure}

We compared the \chandra\ image of the  galaxy center of IC342 with the optical observations.  The registered positions of the X-ray nuclear sources C12 and C13 are overlaid on the \hst\ {\it V\/} band and H$\alpha\/$ images of the central $17''\times17''$ region of the IC342 in Figures~\ref{f:vband}~(top) and~\ref{f:halpha}~(top) respectively.  The spatial extent of the C12 core emission is also shown by a $4''$ circle ({\it dash\/}) in the figures.  The \hst\ {\it I\/} band image was also analyzed but was found to be similar to the {\it V\/} data and thus not shown.  In the \hst\ images, there are many star forming regions at the center of IC342 with obscured emissions and bright knots.  Both \hst\ images show that the optical emission from the central $d \approx 5''$ of IC342 is concentrated in three star clusters aligned in the north-south direction. The star clusters have also been observed previously in the near-IR CO band by~\citet{Boker1999}.  We named the three clusters here as the nuclear star cluster (NSC; middle), SC1 (top), and SC2 (bottom).  The centroid of C12 lies close to that of the optical maximum of the NSC, with an offset of $0.53''$ measured in all the $V$, $I$, and H$\alpha$\/ images.  The observed separations were only slightly larger than the uncertainty of the X-ray bore sight correction of C12 ($0.45''$) in our \chandra\ image.  

Using the near-IR CO band observations, \citet{Boker1999} deduced that the NSC is a relatively young cluster with an age of $10^{6.8-7.8}$ yrs, and a mass of $M_{\rm NSC}\approx6\times10^{6}~M_{\odot}$. These results favored an instantaneous burst model rather than a constant star formation rate model for the NSC, meaning the dominant stellar population are of the same age as the NSC. Also, NSC could have formed as a result of gas flowing to the nucleus initiated by torques from a nuclear stellar bar. The H$\alpha$ image is consistent with this scenario in that it shows gas spiraling around the NSC (see Figure~\ref{f:vband} and~\ref{f:halpha}).
 
The X-ray contours are overlaid on the broadband {\it V\/} and H$\alpha$\/ images of IC342 in Figures~\ref{f:vband}~(bottom) and~\ref{f:halpha}~(bottom) respectively for a comparison of the diffuse X-ray emission with the detailed structure of the optical galaxy center. These contour lines encircle the central $r\sim 5''$ region of IC342 and its structure is found to generally agree with those observed in the optical wavelengths, despite the lower angular resolution of the X-ray data.
Moreover, the X-ray diffused emission is elongated along the north-south axis, which is similar to the alignment of the three optical nuclear star clusters.  These coincidence suggest that the observed X-ray nuclear emission is associated with the ongoing starburst and the three optical star clusters.  However, the newly detected X-ray nuclear source C13 does not coincide with any bright features in the H$\alpha$\/ image, suggesting that it is not associated with the star formation activities at the galaxy center.

\subsubsection{Comparison of X-ray with radio data}
\label{ss:radio}
For comparisons with the X-ray and optical data, we also obtained Very Large Array (VLA) 2 cm (15 GHz) and 6 cm (5 GHz) radio images of IC342. The observations and the data reduction procedures have already been described in Tsai et al. (2006). 
These images have spatial resolutions of $0.3''$ at 6 cm and $0.1''$ at 2 cm, respectively, with a $0.05''$ uncertainty in the absolute position of detected sources. The positions of all the radio sources detected in the 2~cm and 6~cm are marked in Figures~\ref{f:vband}~(top) and~\ref{f:halpha}~(top). The optical images of IC342 presented in \citet{Tsai2006} were not astrometrically registered for comparison with the radio images. However, our  current analysis presents improved astrometry on the X-ray and optical data, allowing accurate identifications of radio counterparts for the X-ray and optical sources. Our revised astrometry explains the slight differences in the radio source positions of our Figures~\ref{f:vband} and~\ref{f:halpha} compared with the Figure~2 of \citet{Tsai2006}.

The brightest radio source, source A using the naming scheme of \citet{Tsai2006}, is near the optical star cluster SC1 and it was identified as a supernova remnant.  One of the radio source, source J, lies very close to the center of the optical emission from NSC at a distance of $0.4''$, and a distance $0.5''$ to the center of the nuclear X-ray source C12. This source is very weak, with only a $1.4\sigma$ detection in 2~cm and was actually not detected in the 6~cm data.  \citet{Tsai2006} explained source J as a \ion{H}{2} region situated between SC1 and NSC, requiring the excitation from a star cluster powered by at least 70 O7 stars.  However, with our corrected astrometry, source J was found to be much more likely to be associated with the center of the NSC. We will discuss the possibility of the radio emission from a radio-quiet AGN in Section~\ref{ss:agn}. It should also be noted that none of the radio sources identified as \ion{H}{2} regions by \citet{Tsai2006}, except source J, could be associated with any star clusters in the optical images.  This could be due to the high internal extinction $A_{\rm V}$ in the central regions of IC342, consistent with the presence of high molecular gas concentrations, which could fuel more star formation in the region.  

The radio contours from the 6 cm VLA image are overlaid on the \hst\ image and \chandra\ contours in Figures~\ref{f:vband} (bottom) and~\ref{f:halpha} (bottom).  The morphology of the radio emission in the central region is very different from both the X-ray and optical images. C12 is encompassed by several radio sources aligned from northeast to southwest that have been identified as supernova remnants and \ion{H}{2} regions~\citep{Tsai2006}. The majority of the bright radio sources and the radio diffused radio emission are located west of C12, suggesting enhanced star forming activity in that region. 

On the other hand, the faint X-ray source C13 is about $3''$ from a compact \ion{H}{2} region (source L in Tsai et al. 2006).  We noticed that the position of the source L listed in Tsai et al. (2006) was incorrect and the revised position should be R.A. = 03h:46m:49.13s, decl. = +68d05m46.4s (Tsai 2007, private communication), which was used here.  With the pointing accuracy of the \chandra\ data at $\approx0.5''$, we concluded that C13 is unlikely to be associated with any radio emissions.

\subsection{Variability of the X-ray nuclear source luminosity}
\label{ss:lum}
To investigate the long term variability of the nuclear source C12, we compared the observed X-ray luminosities at the center of IC342 over a period of 13 years using data taken by \chandra\ (this paper; 2006 April 2), \xmm\ (B03, K03; 2001 February 11), 
and  \rosat\ (\citealt{Bregman1993}; 1991 February 13). We could not directly compare luminosities from the literature because of the different assumptions made in those works.  For example, the luminosity values in the earlier \asca\ and \rosat\ publications were computed assuming a distance to IC342 of 4.5~Mpc.  We recalculated the luminosities from all the observations by assuming the same energy range, spectral model, and assumed distance of IC342.  Thus instead of quoting from the literature, we extracted the fluxes of C12 from the archival data of \xmm\ and \rosat, together with our \chandra\ observation, and analyzed them with the procedures described in Section~\ref{s:anlypho}.  The fluxes from the different observatories were converted from source counts using the spectral model derived from \xmm\ since it had the highest spectral resolution.  We adopted the simpler spectral model assumed for C12 (named X21) in K03, i.e., a power-law model with $N_{\rm H}=(0.32\pm0.05)\times10^{22}$cm$^{-2}$ and $\alpha=2.53\pm0.16$.  Furthermore, we scaled the luminosities and their errors to a distance of 1.8 Mpc and to the same energy range as HRC-I (0.1--10~keV).  The scaled luminosities and errors of the nuclear source C12 from the different X-ray missions are listed Table~\ref{t:lumtable}.  The  results from the two brightest X-ray sources in our \chandra\ image, C3 and C6, are also listed for reference.  The spectral models used were:

\begin{enumerate}
	\item C3: power-law model with $N_{\rm H}=0.60^{+0.06}_{-0.05}\times10^{22}$cm$^{-2}$ and $\alpha = 1.72^{+0.08}_{-0.08}$
	\item C6: disk blackbody model with $N_{\rm H}=1.81^{+0.29}_{-0.25}\times10^{22}$cm$^{-2}$ and $kT=2.00^{+0.24}_{-0.21}\ \rm keV$    
\end{enumerate}

We note that there was an \asca\ observations taken on September 1993. However, the poor spatial resolution of \asca\ (FWHM $>1'$) and the large extraction radius ($2'$) used in analyzing the nuclear source~\citep{Okada1998}~implies that the \asca\ results might suffer from confusion problems. In our \chandra\ data, a $2'$ circular region centering at C12 would have included six other sources (C9, C10, C13, C14, C15, and C18). Using the present \chandra\ photometric information, we estimated that 46\% of the measured \asca\ counts of C12 could be due to confusion, implying a 0.1--10 keV luminosity of $11.8\times10^{38}\lum$. However, this estimate did not include uncertainties due to long term spectral and flux variability information of all these X-ray sources. As a result, we did not include the \asca\ data point in the present variability study. On the other hand, though the spatial resolution of \rosat\ and \xmm\ were lower than \chandra, the confusion problem in the two observations were not as significant as that in \asca. The spatial extent of C12 were similar in both \xmm\ and \rosat, about $8''-10''$, and a circular region of this size would include both C12 and C13.


Nonetheless, the source counts of C13 is only 6\% of C12, and thus its contribution is negligible unless it significant varies in flux. Thus, we decided to ignore the confusion effect in the \xmm\ and the \rosat\ data. 

This analysis is the first to study the luminosity of C12 over a long period of time.  In the energy range 0.1--10 keV, the observed luminosity of the IC342 nuclear X-ray source C12 varies roughly by a factor of 2 over the 3 observations spanning a period of over 15 years.  The observed variations of the luminosity of C12 is actually the smallest amongst the three X-ray sources investigated. It is worth noting that the flux uncertainties tabulated in Table~\ref{t:lumtable} only include photon statistics. Uncertainties in the relative calibration of various X-ray detectors have not been included.  However, previous studies indicate that variations of normalized flux in different X-ray instruments are small, at a level of $\pm10\%$ \citep{Snowden2002}, and thus are not expected to account for all the flux variability observed. Another uncertainty in the current study is the assumption of a single spectral model of C12 over the entire period as there was no spectral information from the \chandra\ and \rosat\ data.

We also studied the short-term variability of C12.  We extracted the source and background lightcurves of C12 using DMEXTRACT and high level background intervals were filtered.  The short-term lightcurves of C12 were studied.  However, we were not able to detect any short-term variability on timescale of the observation because of the observed low count level.

\section{Nature of the Nuclear X-ray Emission}
\label{s:nuclear}

The high resolution \chandra\ HRC-I observation resolves for the first time the X-ray emission at the center of IC342, revealing the structure of the nucleus of this starburst galaxy. As seen from Figure~\ref{f:vband} and~\ref{f:halpha}, both the X-ray and the optical emission are confined to a region of radius $\sim100\rm~pc$ in the center of IC342. The most intense X-ray emission (inner $4''$) is coincident with the NSC seen in both {\it V} band and H$\alpha$ images. There are many regions of star formation within the central region of IC342 which were clearly visible in H$\alpha$. It is this region that contains the 3 optical star clusters and includes our detected \chandra\ X-ray source C12. The extension seen in the X-ray diffuse emission of C12 is most probably due to starburst activities induced by star clusters and massive stars. This region also contains radio sources which aligns from northeast to southwest. As discussed in~\citet{Tsai2006}, these radio sources could be interpreted as \ion{H}{2} regions which were small clusters that required excitation by individual O star clusters, each with over 5000 O stars (\citealt{Becklin1980};~\citealp*{Turner1983}). Possible contributors to the X-ray emission from a nuclear starburst include supernovae, O stars, High Mass X-ray Binaries (HMXBs), ULXs, and an obscured AGN. Using our \chandra\ X-ray data, we attempted to explain in the following the stellar nature in the NSC.

\subsection{Contributions from star formation activities}
Many previous infrared observations of IC342 \citep{Boker1997,Schinnerer2003,Meier2005} have proposed a starburst scenario for the nuclear region. The model of the central starburst includes a nuclear star cluster surrounded by a mini spiral arms-ring system on the scale of $\sim30''$ (e.g., Fig.~10 in~\citealt{Meier2005}), as supported by observations of molecular line emissions.  The X-ray image generally agrees with the spatial scales of this model, with the size of the C12 emission $\sim 15''\times 10''$ is roughly the size of the proposed central ring ($\sim12''$), while the mini spiral arms that extend in the north-south direction to $\sim45''$ would be beyond the spatial extent of C12 (see Figure 1 of~\citealt{Meier2005} for a comparison of the optical and $^{12}$CO~(1-0) infrared emission).  This model suggests that most of the star formation occurs in the central ring region rather than along the spiral arms. Our \chandra\ observation is consistent with this scenario since the X-ray emission is concentrated in the central 100~pc.  This star formation structure is smaller in size compared to those in other starburst galaxies, which have radii of order 1~kpc \citep{Boker1997}.

We could study the starburst properties of IC342 using the results from starburst galaxies and AGN survey~\citet{Mas1995} with the far-infrared to soft X-ray luminosity ratio.  The far-infrared flux measured by the {\it Infrared Astronomical Satellite\/} ($IRAS$) at 60~$\mu$m was $F_{60\mu \rm m} = (260\pm20)$~Jy in the inner r=$16'$ region of IC342~\citep{Beck1988}.~\citet{Tsai2006} found that the luminosity of the compact \ion{H}{2} regions in the nucleus was of the order of $\sim5-10\%$ to the total IR luminosity of IC342.  Therefore we estimated the far-infrared flux in the galaxy center of IC342 to be $F_{60\mu \rm m}(C12)\sim13-26$~Jy.  We derived the soft ($0.5-4.5$~keV) X-ray flux of C12 with WebPIMMS using the spectral model and parameters listed in Section~\ref{s:anlypho} and obtained $L^{\rm C12}_{\rm 0.5-4.5~keV} = (1.9\pm0.3)\times10^{38}\lum$. This led to a flux ratio $\log(L^{\rm C12}_{\rm 0.5-4.5~keV}/L^{\rm C12}_{60\mu \rm m})$ with a range of -3.1 to -3.4. This is in agreement with the value for star formation galaxies, -3.33 ($\sigma=0.47$), and slightly smaller than the average value of Seyfert 2 galaxies, -2.54 ($\sigma=0.81$)~\citep{Mas1995}.

\citet{Ward1988} derived a model to predict X-ray emission from starbursts using the Brackett $\gamma$ emission. The model suggests that the X-ray emission arises from the accretion of material in population I binary systems which are formed together with the OB stars.  The total X-ray emission in the starburst is calculated to be
\begin{equation} 
	L_{0.5-3\rm~keV}=7\times10^{-35}\frac{L_{B\gamma}L_{\rm Xbin}}{R}~(\lum)\ , \ 
\label{eq:LXbin}
\end{equation}
where $L_{\rm B\gamma}$ is the observed Brackett $\gamma$ emission, $L_{\rm Xbin}$ is the average X-ray luminosity of binary systems and was assumed to be $10^{38}\lum$, and $R$ is the number of OB stars per massive binary systems and was assumed to be 500~\citep*{Fabbiano1982}. The Brackett $\gamma$ emission from the inner 100~pc of IC342 was measured to be $\approx6.5\times10^{37}\lum$~\citep{Boker1997}, thus implying a predicted X-ray luminosity from the starburst of $L_{\rm 0.5-3~keV}=9.1\times10^{38}\lum$. Our observed 0.3--5 keV X-ray luminosity of C12 is about 20\% of this predicted luminosity. 
This might suggest that the X-ray emission from the starburst component in C12 was not dominated by population I binary systems as suggested in the~\citet{Ward1988} model. It might instead be due to thermal emission such as coronal from young stars, supernovae/supernova remnants, and the hot interstellar medium heated by supernovae and stellar winds. This might suggest that the soft X-ray emission in C12 was due to part of starburst component and some other thermal emission. We will discuss the possible thermal origin of the diffuse emission from C12 in section~\ref{ss:other}.  

\subsection{Contributions from AGN}
\label{ss:agn}

Another possible origin of the core X-ray emission from the galactic center is an AGN. Using the \xmm\ data,  B03 proposed that the hard X-ray from C12 could be coming from a hidden AGN, although there was no direct evidence to confirm its existence in IC342. The starburst-AGN connection has been a topic that generated great interests but the exact relationship remains uncertain. The triggering mechanism for both phenomena could be the interacting or the merging of gas-rich galaxies, which generates fast compression of the available gas in the inner galactic regions, causing both the onset of a major starburst and the fueling of a central black hole, hence raising the AGN. Examples of this class of objects include NGC6240~\citep{lira2002}, NGC4303~\citep{Jimenez2003}, and NGC1808~\citep{Jimenez2005}. One particularly interesting object for comparison is NGC1808, where a similar core+diffuse X-ray emission morphology was observed. The extended X-ray emission of the NGC1808 center ($\sim$850~pc) is dominated by soft radiation ($\le$ 1 keV), and follows the same orientation as the H$\alpha$ and optical-UV emission, suggesting intense star formation in the region. In contrast to IC342, two X-ray point sources (with a separation of $\sim65$~pc) were resolved in the core of NGC1808. Spectral information from the \chandra\ ACIS data showed that they were associated respectively with hot gas emission and hard X-ray emission. The nuclear location from the 2MASS data of NGC1808 was closer to the softer point source (source S1 in~\citealt{Jimenez2005}). Furthermore, both the X-ray spectra of the centers of IC342 and NGC1808 can be modeled with absorbed power-law plus MEKAL models, and the Fe K$\alpha$ emission is missing in both spectra. \citet{Jimenez2005} compared the ratios of the detected lines such as \ion{O}{8} and \ion{Ne}{9} in NGC1808 with those observed in M82 and concluded a starburst-AGN coexistence in NGC1808.


With astrometrically corrected X-ray and radio data, we were able to study the radio loudness of the possible AGN at the center of IC342. As discussed in Section~\ref{ss:radio}, radio observations of IC342 reveal a group of weak sources within 5$''$ radius of the center, with an integrated $L_{\rm 6cm}\approx10^{35.6}\lum$. This is slightly lower than the the range of $10^{36}-10^{39}\lum$ observed from the 48 low-luminosity AGNs (LLAGNs) sampled by \citet{Terashima2003}. A weak radio source, source J, is near the X-ray center source C12, with $F_{\rm 2cm}=0.7\pm0.5$~mJy, and $F_{\rm 6cm}<0.2$~mJy (3-$\sigma$ limit), while a stronger source, source A, with $F_{\rm 2cm}=0.9\pm0.3$~mJy, and $F_{\rm 6cm}=1.5\pm0.2$~mJy, is $0.99''$  away from C12 \citep{Tsai2006}. Therefore if we concentrate on the central $2''$ radius region of IC342, corresponding to the size of the core component of C12, then the integrated radio flux in the region would be $\sim1\times10^{-16}\flux$. Together with the hard (2~-- 10~keV) X-ray flux of the C12 core determined from WebPIMMS ($F_{\rm 2-10~keV}^{\rm C12 core} = 9.1\times10^{-14}\flux$), we could determine the radio loudness parameter $R_{\rm X}=\frac{L_{\rm 6cm}}{L_{\rm 2-10~keV}}$, as defined by~\citet{Terashima2003}, to be $\log R_{\rm X}\approx-3$. This value is comparable to the average for Seyfert galaxies ($-3.3^{+0.3}_{-1.0}$) as determined from a sample of 51 AGN candidates \citep{Capetti2006}. 

Based on the distribution and flux of the radio sources in the nucleus, we could classify that IC342 as a radio-quiet type AGN, if there existed one. Radio-quiet AGN included LINERs, Seyfert (1 and 2) galaxies, and QSOs~\citep{Capetti2006}. 
Assuming the X-ray luminosity of C12 were due entirely to an AGN, then the hard X-ray luminosity observed  $L_{\rm 2-10~keV}^{\rm C12}=(5.4\pm0.5)\times10^{37}\lum$ would be just below the lower range of other LINERs observed with \rosat, \asca, and \chandra\ (in the range $10^{38}-10^{41}\lum$), but would be low when compared to Seyfert 1 galaxies or quasars (typically $>10^{43}\lum$). Another property to consider is the X-ray flux variability. As shown in Table~\ref{t:lumtable}, we observed the IC342 center X-ray flux to have varied by a factor of $\sim2$ (or flux variability amplitude of $\sim60\%$) over 15 years in the three observations, which is comparable to the 
mean long term flux variability amplitude of 68\% for type~1 AGN and 48\% for type~2 AGN~\citep{Beckmann2007}. 

On the other hand, we also notice that the power law index of the X-ray central source determined from the \xmm\ spectral fit, $\alpha=2.53\pm0.16$ (K03), is slightly steeper than the typical value of radio-quiet AGN and LINERs at $\alpha=1.7-2.3$ \citep{Reeves2000,George2000,Georgantopoulos2002}. Moreover, the absorption column density, $N_{\rm H}\sim10^{22}\rm cm^{-2}$ (B03, K03), is much lower than typical AGN values of $N_{\rm H}\geq10^{24}\rm cm^{-2}$, together with no intrinsic absorption in the X-ray spectral fit. 
There was also no sign of the 6.4~keV Fe~K$\alpha$ line emission, typically an indicator of AGN, in the \xmm\ EPIC spectrum (B03, K03). However, this line is also absent in a number of starburst galaxies identified as having hidden AGN~\citep{Tzanavaris2007} and thus we cannot exclude the possibility of the presence of an AGN in IC342. 

The positional coincidence of the X-ray center with the optical star clusters in IC342 could suggest a scenario of coexistence of the nuclear star cluster and an AGN, as discussed by other authors~\citep{Seth2008,Shields2008}. We attempt to place a lower limit to the black hole mass. Assuming that the black hole is accreting at sub-Eddington rate, we could place a lower limit of the black hole mass at the center of IC342 at $M_{\rm BH}\ge3\times10^{4}M_{\odot}$ given the bolometric luminosity of the galaxy $L_{\rm bol}\approx3\times10^{42}\lum$ as determined by \citet{Becklin1980}. On the other hand, it had been reported that nuclear star clusters are more massive than the coexisting black holes and their mass ratio $M_{\rm BH}$/$M_{\rm NSC} \sim 0.1 - 1$~\citep{Seth2008}. This implies $M_{\rm BH}\le6\times10^{6}M_{\odot} = M_{\rm NSC}$, the nuclear star cluster mass as determined by IR data by~\citet{Boker1999}. Another estimate of the black hole mass comes from the $M_{\rm BH}-\sigma_{\rm bulge}$ relation~\citep{Gebhardt2000,Tremaine2002,Hu2008}, which had been shown to apply even in the massive star cluster regime~\citep{Gebhardt2005,Shields2008}. Using the velocity dispersion measurements for the nuclear star cluster ($\sigma_{\rm NSC}=33$km~s$^{-1}$,  \citealt{Boker1999}), and the $M_{\rm BH}-\sigma_{\rm bulge}$ relation of~\citet{Gebhardt2000}, we derived $M_{\rm BH}\approx1\times10^{5}M_{\odot}$, which is consistent with the limits described above. Based on these calculations, the range of black hole mass is suggestive of an intermediate mass black hole (IMBH). However, we should not rule out the alternative possibility of an inactive supermassive black hole (mass $\sim10^{6}-10^{7}M_{\odot}$) similar to known Seyfert nuclei that the nuclear emission is due to advection-dominated accretion flows (ADAF;~\citealt{Rees1982,Narayan1995,Fabian1995}). Further spectral analysis could help to address whether this is the case.

\subsection{Other X-ray contributions associated with star formation}
\label{ss:other}
The spatial extent of the X-ray emission from the nucleus of IC342 could alternatively indicate the presence of populations of stellar objects uniformly distributed throughout the core region. These objects could be X-ray binaries, OB stars, or supernova remnants. For example, as pointed out by~\citet{Stevens1999}, populations of massive X-ray binaries evolved from high mass stars could be formed in starburst events, and such populations would be consistent with models associated with starbursts. It had been suggested that photoionization from these stars might have accounted for the the enhanced $C^{+}$ millimeter emission in the central ring of the mini spiral model of IC342 \citep{Meier2005}.


We considered the possibility of multiple unresolved point sources, consisting of a mixture of stars and XRBs in the nuclear star cluster, contributing to the observed X-ray structure at the center of IC342. A burst of star formation usually resulted in the production of large number of OB stars \citep{Stevens1999}. With the typical X-ray luminosity for O stars $\approx10^{31}-10^{32}\lum$~\citep{Sciortino1990}, and the number of O stars in the nucleus of IC342 at $\sim$4000 estimated from the Brackett $\gamma$ emission~\citep{Ward1988}, we derived the X-ray luminosity from these stars at $\log L_{\rm X}\approx 34.6-35.6$. This is less than 0.1\% of the total observed X-ray luminosity of C12. The more luminous HMXBs, usually associated with young stellar population like O stars, might have also contributed to the observed X-ray emission. This could be inferred from the ratio of total X-ray luminosity from HMXB (with $L_{\rm X}\le10^{37}\lum$) to the number of O stars, i.e. $(2-20)\times10^{34}$~\citep{Helfand2001}. For the nucleus of IC342, $L_{\rm Xbin}$ was estimated to be about $8\times10^{37-38} \lum$ (10\%--90\% of X-ray luminosity of C12), and the corresponding number of X-ray binaries was about 10 as inferred from $L_{\rm B\gamma}$. If these O stars and XRBs were uniformly distributed in a region of  radius 50 pc (the central $4''$ region of C12), the mean separation ($\approx0.5''$ for HMXBs and $\le0.01''$ for O stars) would be smaller than the spatial resolution of \chandra\ HRC-I. As a result, a model with 4000 O stars and $\approx10$ HMXBs could account for both the observed morphology and flux of core component of the X-ray emission of C12. 

We also considered whether the diffuse component of C12 could also be explained by the model above. As shown in Figure~\ref{f:halpha}, the nucleus contains many regions of star formation, which could be the result of an outflow induced by the stellar winds of the massive stars or X-ray binaries present in the star clusters, as observed in M82 \citep*{Strickland1997,Moran1997}. The exact nature of this diffuse emission remains uncertain given the limited amount of X-ray data available. Further observations with possible spectral analysis of this diffuse emission could help to link the starburst activities and the X-ray emission of the stellar objects in the nuclear star cluster.

\section{SUMMARY AND CONCLUSION}
\label{s:summary}
This paper aimed to study the nature of the X-ray emission in the nucleus of IC342 using high resolution ($\theta_{\rm FWMH}\approx0.5''$) \chandra\ HRC-I observation. The historical ULX at the center of IC342, X3, was resolved into 2 sources, namely C12 and C13, for the first time since the  X-ray telescopes prior to \chandra\ did not have enough spatial resolution. The brighter source C12, with over 90\% of the X-ray nuclear flux, was found to be located very close to the optical galactic center. The source C12 was also revealed as an extended source with a complex X-ray morphology, with a core component of diameter $4''$ that is roughly consistent with a point source of $L_{0.08-10\rm~keV}=(6.7\pm0.5)\times10^{38}\lum$, and a diffuse component that extends out to a diameter of $\sim10''$, or $\sim150$~pc. Thus, we concluded that C12 is not an ULX. The much weaker source C13, with $L_{0.08-10\rm~keV}=(5.0\pm1.4)\times10^{37}\lum$, is consistent with a point source situated $6.51''$ at P.A. = $240^{\circ}$ from C12.

Registration of the X-ray image with optical (\hst) and radio (VLA) data allowed us to find the counterparts of X-ray sources. The registered position of C12 is R.A. = 03h:46m:48.43s, decl. = +68d05m47.45s, which is located $0.53''$ from the optical NSC ~\citep{Boker1999} and $0.50''$ from a weak radio source (source J in~\citealt{Tsai2006}), and within the boresight correction $0.45''$ of the \chandra\ data. Intense star formation was observed in the \hst\ H$\alpha$ image of the inner 100 pc of IC342, where the 3 optical star clusters (SC1, SC2, and NSC) and our \chandra\ X-ray source C12 are found. Moreover, the diffuse X-ray emission extends preferentially along the north-south direction, which is the same as the alignment of the star clusters in the optical. This suggests that the nuclear X-ray emission is associated with star formation. 

We discussed possible contributors to the X-ray emission in the nuclear region of IC342, with a main focus on the extended source C12. A starburst model for IC342 based on molecular line observations~\citep{Meier2005} that included a mini spiral arms-ring system was found to be consistent with the spatial extent of C12. Furthermore, the predicted X-ray emission from starbursts as derived from the Brackett $\gamma$ emission \citep{Ward1988} is close to the actual observed X-ray luminosity of C12, indicating that X-ray from the nucleus is dominated by starburst activities. On the other hand, the hard emission from previous \xmm\ observation suggested that there could be an AGN in IC342 (B03, K03). The luminosity of C12, $L_{\rm 0.08-10~keV}=(6.7\pm0.5)\times10^{38}\lum$, is consistent with that of a LINER at (typical luminosity $10^{38-41}\lum$). By comparing our current observation with earlier X-ray missions, we found a possible long term variability factor of 1.8 in the IC342 nuclear X-ray emission over 15 years. This amount of variation would be consistent with that from an AGN. Weak radio emission of $L_{\rm 6cm}\sim10^{35}\lum$ in the vicinity of C12 is consistent with a radio-quiet AGN. The observed properties (luminosity, absorption column density) are similar to those of type 2 low-luminosity LINERs or Seyfert 2 galaxies. The black hole mass as inferred from the $M_{\rm BH}-\sigma_{\rm bulge}$ relation for the AGN of $\approx1\times10^{5}M_{\odot}$, is consistent with an IMBH, but we should not ignore the possibility of an inactive supermassive black hole. 
In conclusion, our current data did not confirm the existence of an AGN in the core of IC342, but it could neither be discarded.

The lack of any spectral information of our \chandra\ HRC-I data did not allow for detailed studies of the surface brightness distribution of the nuclear source C12 in different X-ray energy bands. In particular, separate spectral information on the core and diffuse components of the X-ray galactic center emission of IC342 would allow us the opportunity to put a tighter constraint on the co-existence of a low luminosity AGN and a starburst in IC342. With the X-ray emission from the core and diffused components at roughly the same level, further high spectral resolution  observations with the \chandra\ ACIS could shed new light on how the nucleus of IC342 fit into the starburst-AGN family.

\section{ACKNOWLEDGEMENTS}

We thank L. Sjouwerman and C.W. Tsai for generously providing the radio maps and helpful discussions. This work is based on observations obtained with \chandra, support for this work was provided by the National Aeronautics and Space Administration through Chandra Award Number G06-7110X issued by the Chandra X-ray Observatory Center, which is operated by the Smithsonian Astrophysical Observatory for and on behalf of the National Aeronautics Space Administration under contract NAS8-03060.  S.Y.~Mak acknowledges support from HKU under the grant of Postgraduate Studentship. C.S.J.~Pun acknowledges support of a RGC grant from the government of the Hong Kong SAR. A.K.H.~Kong acknowledges support from the National Science Council, Taiwan, through a grant NSC96-2112-M-007-037-MY3.


\begin{deluxetable}{ccccccccl}
\tablecolumns{9}
\tabletypesize{\scriptsize}
\tablecaption{\chandra\ HRC source list of IC342}
\tablehead{
\colhead{ID} &
\colhead{RA} & \colhead{DEC} & \colhead{RA} & \colhead{DEC} & \colhead{Net Counts} & \colhead{S/N} & \colhead{$L_{X}(0.08-10$ keV)} & \colhead{Remark} \\
\colhead{} & \colhead{J2000.0} & \colhead{J2000.0} & \colhead{error ($''$)} & \colhead{error ($''$)} & \colhead{} & \colhead{} & 
\colhead{(10$^{38}$ \lum)} & \colhead{} \\
\colhead{(1)} & \colhead{(2)} & \colhead{(3)} & \colhead{(4)} & \colhead{(5)} & \colhead{(6)} & \colhead{(7)} & \colhead{(8)} & \colhead{(9)}
}
\startdata
C1&3:45:10.62&68:02:31.95&1.1&0.7&$83.32\pm14.89$&5.60&$2.81\pm0.50$& \\
C2&3:45:40.52&68:03:10.65&0.7&0.5&$59.19\pm12.02$&4.93&$1.99\pm0.41$& \rosat\#1, XMM X2 \\
C3&3:45:55.62&68:04:55.89&0.2&0.4&$794.40\pm29.24$&27.17&$26.75\pm0.98$& IC342 X1, \rosat\#3, XMM X7 \\
C4&3:45:59.65&68:05:38.62&0.7&0.5&$36.68\pm9.64$&3.81&$1.24\pm0.32$& XMM X9\\
C5&3:46:06.58&68:07:05.70&0.5&0.4&$82.71\pm11.59$&7.14&$2.79\pm0.39$& \rosat\#4, XMM X12, $B$=12.8 $R$=11.82 \\
C6&3:46:15.79&68:11:12.98&0.3&0.4&$404.79\pm21.57$&18.76&$13.63\pm0.73$&IC342 X2, \rosat\#5, XMM X13 \\
C7&3:46:22.17&68:05:05.84&0.4&0.4&$19.03\pm6.29$&3.03&$0.64\pm0.21$& XMM X14 \\
C8&3:46:27.36&68:04:10.83&0.5&0.4&$17.10\pm4.52$&3.79&$0.58\pm0.15$& \\
C9&3:46:43.66&68:06:11.58&0.2&0.4&$64.10\pm8.44$&7.60&$2.16\pm0.28$& \rosat\#6, XMM X19 \\
C10&3:46:44.95&68:05:55.00&0.2&0.4&$10.86\pm4.41$&2.46&$0.37\pm0.15$& \\
C11&3:46:45.56&68:09:46.43&0.9&0.4&$16.46\pm5.32$&3.09&$0.55\pm0.18$& \rosat\#7, XMM X20 \\
C12&3:46:48.43&68:05:47.45&0.2&0.4&$254.60\pm17.37$&14.66&$8.57\pm0.59$& nucleus, IC342 X3, \rosat\#8, XMM X21 \\
C13&3:46:49.50&68:05:45.10&0.3&0.4&$15.08\pm4.00$&3.77&$0.51\pm0.14$& \\
C14&3:46:52.82&68:05:04.96&0.2&0.4&$8.47\pm3.54$&2.40&$0.29\pm0.12$& XMM X24\\
C15&3:46:57.49&68:06:19.51&0.2&0.4&$108.83\pm11.13$&9.78&$3.66\pm0.38$& \rosat\#9, XMM X26 \\
C16&3:47:01.86&68:02:37.02&0.6&0.5&$24.24\pm6.50$&3.73&$0.82\pm0.22$& \\
C17&3:47:04.15&68:09:05.82&0.5&0.5&$15.49\pm5.35$&2.89&$0.52\pm0.17$& \\
C18&3:47:04.89&68:05:18.18&0.2&0.4&$11.02\pm3.64$&3.02&$0.37\pm0.12$& \\
C19&3:47:10.90&68:02:53.00&0.5&0.5&$14.70\pm5.14$&2.86&$0.50\pm0.17$& \\
C20&3:47:18.38&67:51:14.59&1.4&0.6&$307.63\pm34.62$&8.88&$10.40\pm1.20$& {\it B}=11.7 {\it R}=10.0 \\
C21&3:47:18.73&68:11:29.79&0.8&0.5&$88.41\pm12.01$&7.36&$2.98\pm0.40$& XMM X30 \\
C22&3:47:22.91&68:08:59.81&0.4&0.4&$72.08\pm9.62$&7.49&$2.43\pm0.32$& XMM X31 \\
C23&3:48:06.93&68:04:54.32&0.8&0.7&$28.70\pm8.10$&3.54&$0.97\pm0.27$& \\
\enddata
\tablecomments{
Column (1) lists the running ID number of the X-ray sources. "C" here stands for \chandra. 
Column (2) and (3) list the X-ray source positions using the astrometric reference as discussed in Section~\ref{s:anlyast}.
Column (4) and (5) list the errors of RA and DEC in arc seconds, including the errors returned by the WAVDETECT task and the shift used for correcting the astrometry.
Column (6) lists the the background subtracted counts and the errors, as discussed in Section~\ref{s:anlypho}.
Column (7) lists the signal to noise ratios of X-ray sources.
Column (8) lists the unabsorbed 0.08-10 keV luminosities and their errors in unit of $10^{38}\lum$, assuming the average power-law spectrum of $N_{rm H} = 8\times10^{21}$~cm$^{-2}$ and $\alpha =2$ as derived by K03, and a distance to IC342 of 1.8~Mpc. 
Column (9) indicates the counterparts of the X-ray sources. The 3 brightest X-ray sources, i.e. X1, X2, and X3 are marked~\citep*{Fabbiano1987}. X-ray sources associated with stars are indicated by their {\it B} and {\it R} magnitudes taken from the USNO-B1.0 Catalog.}
\label{t:sourcelist}
\end{deluxetable}

\begin{deluxetable}{cccc}
\tablecolumns{4}
\tablewidth{0pc}
\tabletypesize{\scriptsize}
\tablecaption{Comparison of luminosities of the 3 brightest X-ray sources in IC342 in different observations}
\tablehead{
\colhead{} &          \multicolumn{3}{c}{Luminosity in 0.1-10 keV (10$^{38}$ \lum)} \\
\colhead{Source} & \colhead{\chandra} & \colhead{\xmm}  & \colhead{\rosat} \\
\colhead{} & \colhead{(2006 April 2)} & \colhead{(2001 February 11)} & \colhead{(1991 February 13)} \\
\colhead{(1)} & \colhead{(2)} & \colhead{(3)} & \colhead{(4)}  
}
\startdata
C3&$18.6\pm0.7$&$19.9\pm0.5$&$10.3\pm0.1$\\
C6&$25.7\pm1.4$&$18.7\pm1.9$&$50.5\pm10.5$\\
C12&$5.7\pm0.4$&$10.3\pm0.7$&$7.1\pm0.8$
\enddata
\tablecomments{ 
Column (1) lists the 3 brightest X-ray sources in IC342: C3, C6, and C12.
Columns (2) to (4) list the unabsorbed luminosities and 1-$\sigma$ uncertainties of the X-ray sources in observations by \chandra\ (this paper), \xmm, and \rosat.
The uncertainties include errors in count rate and errors in spectral model parameters. They have been scaled to the same distance of 1.8~Mpc, the same energy range (0.1-10 keV), and the same spectral model~\citep[Table 2]{Kong2003} (cf. parameters listed in Section~\ref{ss:lum}).\\
} 
\label{t:lumtable}
\end{deluxetable}

\end{document}